\documentclass[hidelinks,journal,10pt]{IEEEtran}
\usepackage{amsmath,amsfonts,amssymb}
\usepackage{algorithm, algpseudocode}
\usepackage{array}
\usepackage[font=small]{subfig}
\usepackage{textcomp}
\usepackage{url}
\usepackage{verbatim}
\usepackage{graphicx}
\usepackage{cite}
\usepackage{orcidlink}
\usepackage[acronym,shortcuts]{glossaries}
\usepackage[sans]{dsfont}
\usepackage{bm}
\usepackage{relsize}
\usepackage{soul}
\usepackage{outlines}
\usepackage{dblfloatfix}
\usepackage{lipsum}
\usepackage[bottom]{footmisc} 

\newcommand{\trans}[0]{^{\mathsf{T}}}
\newcommand{\transs}[0]{^{\!\mathsf{T}}}

\newcommand{\herm}[0]{^{\mathsf{H}}}
\newcommand{\R}[0]{^{\mathrm{R}}}
\newcommand{\x}[0]{^{\mathsf{x}}}
\newcommand{\cp}[0]{^{\mathrm{c}}}
\newcommand{\I}[0]{^{\mathrm{I}}}
\newcommand{\diag}[1]{{\mathrm{diag}}\!\left({#1}\right)}

\newcommand{\Exp}[0]{\mathbb{E}}

\newcommand{\Real}[1]{\Re\{{#1}\}}
\newcommand{\Imag}[1]{\Im\{{#1}\}}

\newcommand{\srvR}[0]{\hat{\mathbf{e}}_{k^{\mathrm{R}}_p:n}}
\newcommand{\srvI}[0]{\hat{\mathbf{e}}_{k^{\mathrm{I}}_p:n}}
\newcommand{\srvRp}[0]{\hat{\mathbf{e}}_{k^{\mathrm{\!\;\!R}}}{}_{\!\!\!\!{}_{p'}\!:n}}
\newcommand{\srvIp}[0]{\hat{\mathbf{e}}_{k^{\mathrm{I}}}{}_{\!\!\!{}_{p'}\!:n}}
\newcommand{\avRp}[0]{{\mathbf{e}}_{k^{\mathrm{\!\;\!R}}}{}_{\!\!\!\!{}_{p'}}}
\newcommand{\avIp}[0]{{\mathbf{e}}_{k^{\mathrm{I}}}{}_{\!\!\!{}_{p'}}}

\newacronym{HDSIM}{HDSIM}{hyper-dimensional sparse index modulation}
\newacronym{IM}{IM}{index modulation}
\newacronym{SS}{SS}{spread spectrum}
\newacronym{QC}{QC}{quantum computing}
\newacronym{OS-QSM}{OS-QSM}{optimized scalable quadrature spatial modulation}
\newacronym{GSM}{GSM}{generalized spatial modulation}
\newacronym{mMIMO}{mMIMO}{massive multiple-input multiple-output}
\newacronym{MIMO}{MIMO}{multiple-input multiple-output}
\newacronym{MU}{MU}{multi-user}
\newacronym{CF-MIMO}{CF-MIMO}{cell-free MIMO}
\newacronym{MSE}{MSE}{mean squared error}
\newacronym{FG}{FG}{factor graph}
\newacronym{QSM}{QSM}{quadrature spatial modulation}
\newacronym{GQSM}{GQSM}{generalized quadrature spatial modulation}
\newacronym{BP}{BP}{belief propagation}
\newacronym{GaBP}{GaBP}{Gaussian belief propagation}
\newacronym{SM}{SM}{spatial modulation}
\newacronym{6G}{6G}{sixth generation}
\newacronym{IQ}{IQ}{in-phase and quadrature}
\newacronym{ML}{ML}{maximum-likelihood}
\newacronym{BER}{BER}{bit error rate}
\newacronym{P2P}{P2P}{point-to-point}
\newacronym{STC-GSM}{STC-GSM}{space-time coded generalized spatial modulation}
\newacronym{AWGN}{AWGN}{additive white Gaussian noise}
\newacronym{CSI}{CSI}{channel state information}
\newacronym{MQAM}{$M$-QAM}{$M$-ary quadrature amplitude modulation}
\newacronym{IC}{IC}{interference cancellation}
\newacronym{SGA}{SGA}{scalar Gaussian approximation}
\newacronym{CLT}{CLT}{central limit theorem}
\newacronym{MFB}{MFB}{matched filter bound}
\newacronym{PDF}{PDF}{probability density function}
\newacronym{GB-ISTA}{GB-ISTA}{greedy boxed iterative soft-thresholding algorithm}
\newacronym{MP}{MP}{message passing}
\newacronym{SNR}{SNR}{signal-to-noise ratio}
\newacronym{EbN0}{$E_b/N_0$}{energy-per-bit-to-noise-power-spectral-density ratio}
\newacronym{SD}{SD}{sphere decoder}
\newacronym{QAM}{QAM}{quadrature amplitude modulation}
\newacronym{PMF}{PMF}{probability mass function}
\newacronym{ISAC}{ISAC}{integrated sensing and communications}
\newacronym{STC}{STC}{space-time coding}
\newacronym{RIS}{RIS}{reconfigurable intelligent surface}
\newacronym{SotA}{SotA}{state-of-the-art}
\newacronym{IER}{IER}{index vector error rate}
\newacronym{5G}{5G}{fifth generation}
\newacronym{B5G}{B5G}{beyond fifth generation}
\newacronym{mmWave}{mmWave}{millimeter-wave}
\newacronym{THz}{THz}{Terahertz}
\newacronym{RF}{RF}{radio frequency}
\newacronym{STBC}{STBC}{space-time block code}
\newacronym{MMSE}{MMSE}{minimum mean-squared-error}
\newacronym{CS}{CS}{compressive sensing}
\newacronym{i.i.d.}{i.i.d.}{independent and identically distributed}
\newacronym{UVD}{UVD}{unit vector decomposition}
\newacronym{PAM}{PAM}{pulse amplitude modulation}
\newacronym{FLOP}{FLOP}{floating point operation}
\newacronym{VGaBP}{VGaBP}{vector-valued GaBP}
\newacronym{OMP}{OMP}{orthogonal matching pursuit}
\newacronym{OFDM}{OFDM}{orthogonal frequency division multiplex}

\begin{document}
\title{\vspace{-0.5ex}\color{black} Enabling Massive Index Modulation Systems \\[-0.2ex] via Combinatorics-Free Detection}

\author{Hyeon Seok Rou,~{\color{black}\IEEEmembership{Member,~IEEE,}}
Giuseppe Thadeu Freitas de Abreu,~\IEEEmembership{Senior Member,~IEEE,} \\
Takumi Takahashi,~\IEEEmembership{Member,~IEEE,}
David Gonz{\'a}lez G.,~\IEEEmembership{Senior Member,~IEEE,}
and Osvaldo Gonsa.
\thanks{\hspace{-3ex}  H.~S.~Rou and G.~T.~F.~de~Abreu are with the School of Computer Science and Engineering, Constructor University, Campus Ring 1, 28759, Bremen, Germany (e-mails: [hrou, gabreu]@constructor.university).}
\thanks{\hspace{-3ex} T. Takahashi is with the Graduate School of Engineering, Osaka University, Suita 565-0871, Japan (e-mail: takahashi@comm.eng.osaka-u.ac.jp).}
\thanks{\hspace{-3ex} D.~Gonz{\'a}lez~G. and O.~Gonsa are with Continental Automotive Technologies GmbH, Guerickestrasse 7, 60488, Frankfurt am Main, Germany (e-mails: david.gonzalez.g@ieee.org, osvaldo.gonsa@continental-corporation.com).}
\thanks{\hspace{-3.5ex} \textit{A part of this article has been presented at the IEEE International Workshop on Computational Advances in Multi-Sensor Adaptive Processing, 2023} \cite{Rou_CAMSAP23}.}%

\vspace{-4ex}
}

\markboth{To be Submitted to the Transactions on Wireless Communications}%
{}


\maketitle


\begin{abstract}
\color{black}
%
\Ac{IM} is one of the key enabling technologies for \ac{B5G} and \ac{6G} wireless systems, attracting attention for its inherent energy and spectral efficiency resulting from conveying information through the indexation of the resources utilized in during signal transmission.
However, a remaining critical bottleneck for large-scale \ac{IM} is the consequently infeasible detection complexity of combinatoric order.
Therefore in this article, in order to maximally reap the advantages of \ac{IM} in large scenarios, we propose a novel \ac{MP} decoder designed under the \ac{GaBP} framework exploiting a novel \ac{UVD} of \ac{IM} signals with purpose-derived novel probability distributions.
The proposed method enjoys a low decoding complexity that is independent of previously prohibitive combinatorial factors, while still approaching the performance of unfeasible \ac{SotA} search-based methods.
The effectiveness of the proposed approach is demonstrated via complexity analysis and numerical results for the exemplary piloted \ac{GQSM} systems of truly massive sizes (up to 96 antennas).
\end{abstract}

\vspace{-0.5ex}


\glsresetall

$~$
\vspace{-4.75ex}
\section{\color{black}Introduction}
\label{sec:introduction}
\vspace{-0.5ex}

The expectations of \ac{B5G} and \ac{6G} wireless systems prospect significant improvements over the existing systems in various different aspects including reliability, latency, energy and spectral efficiencies, coverage, user capacity, and more \cite{Rappaport_Access19, Vo_MNA22, Uusitalo_Access21, Wang_CST23}, motivating the investigation of a plethora of novel technologies such as \ac{mmWave}/\ac{THz} communications \cite{Tripathi_6G21, Song_TTHz22}, \ac{mMIMO} and \ac{CF-MIMO} \cite{Huo_Sensors23, He_JCIN21}, \acp{RIS} \cite{Zhang_Springer21, Pan_CM21}, and \ac{ISAC} \cite{Liu_JSC22, WangITJ2022, Wei_ITJ23, Rou_SPM24}.

Among these many novel technologies, \ac{IM} \cite{Mandloi_B5G21, Mao_CST19,Ishikawa_CST18,Cheng_WC18, Sugiura_Access17} has recently gained great attention as a potential enabling technology in its inherently high throughput, energy and spectral efficiencies.
Specifically, this attractive and unique feature of enabling frugal utilization of resources at the transmitter is enabled by the activation of only small subsets of the totally available resources to encode information.
This characteristic also lends \ac{IM} schemes great flexibility, making it transparent to other promising technologies, as demonstrated by a quickly expanding literature on the integration of \ac{IM} with classic communications concepts such as \ac{OFDM} \cite{Li_WC20, Wen_TSP15, Wen_TC17, Li_TWC22}, \ac{MIMO} \cite{Mesleh_TVT08, Younis_Asilomar10, Wen_JSAC19, Rou_TWC22}, and multiple access \cite{Raafat_TWC20, Li_Network23}, and next-generation technologies such as \ac{RIS} and \ac{ISAC} \cite{Sui_TC24,Ma_JSTSP21,Gopi_TWC21,Rou_Asilomar24}.

Generally, an \ac{IM} scheme activating $P$ out of a total of $N$ resources can convey up to $\lfloor\log_2\!\binom{N}{P}\rfloor$ additional information bits on top of those conveyed over the selected $P$ resources themselves, for example, via digital symbol modulation.
Therefore, even with sufficiently large $N$ and an adequate $P$, significant amounts of information can be encoded in the index domain, such to highlight the potential of \ac{IM} to outperform conventional systems in certain scenarios \cite{Shamasundar_TWC22, Wen_TSP15}, with the added bonus of improving the energy-efficiency and freeing the unused resources to other functionalities, such as opportunistic communications and sensing \cite{Ma_TVT21, ElMai_CAMSAP23}, etc.

The price of the aforementioned advantages of \ac{IM} are, however, the burden cast onto the receiver in terms of the computational complexity required to detect the activated resources at each given transmission instance, given all possible resource activation patterns in the codebook of size $2^{\lfloor\log_2\binom{N}{P}\rfloor}$.
This challenge has therefore motivated work on low-complexity decoders for \ac{IM}, in its many variations.
To cite some examples, improvements over the brute-force \ac{ML} search based on reduced-search methods such as lattice and sphere decoders \cite{Hu_PC19}, have decreased the total decoding complexity in proportion to the total codebook size, also by exploiting the specific \ac{IM} design properties \cite{Wang_TVT20, Wen_TC17, Li_TWC22}.
However, such methods are still generally dependent on the combinatorial factor, prohibitive for truly massive scales.

More recently, methods aiming to directly reduce the order of the binomial coefficient have also been proposed.
In \cite{An_TVT22}, for instance, the complexity order is reduced from $\binom{N}{P}$ to $\binom{\alpha N}{P}$ with $\frac{P}{N} \leq \alpha \leq 1$, by leveraging the \ac{OMP} algorithm to operate over the probable non-zero indices of the sparse signal of the \ac{GQSM}, which is a specific type of \ac{IM}.
It was shown thereby that a trade-off between computational complexity and optimal performance can be achieved by adjusting the value of $\alpha$.
On the other hand, in \cite{Rou_Asilomar22_QSM}, a \ac{GQSM} detector leveraging the \ac{MP} algorithm with vector-valued variables under the \ac{GaBP} framework is proposed, which reduces decoding cost to the order of the square root of the binomial coefficient, \textit{i.e.}, $\binom{N}{P} {\!\!~}^{\!\frac{1}{2}}$, while approaching the performance of \ac{ML} solution.
Finally, there exists also non-classical approaches which are independent of the binomial coefficient, including techniques based on machine learning (ML) \cite{Katla_Access20, Kim_WCL21} and \ac{QC} \cite{Yukiyoshi_VTC24}.
Such alternatives have, nevertheless, their own limitations, such as the need for exhaustive off-line training or expensive and uncommon quantum computers.


$~$

In light of all the above, we propose in this article a novel framework for \ac{IM} detection which can enable a significantly reduced detection complexity that is complete independent of the prohibitive combinatorial factor inherent to \ac{IM}, while still incorporating the elaborate combinatorial index codebook correlations.
This is enabled by a novel \ac{UVD} representation of \ac{IM} signals{\color{black}, which reduce the effective signal space from the total signal vector variable into smaller unit vectors and scalars with significantly smaller effective signal spaces.} 
Then, further aided by a novel statistical analysis of the non-trivial index-wide activation probabilities {\color{black}of the decoupled variables, tailored \ac{GaBP}-based \ac{MP} algorithms are designed} to enable the concurrent estimation of the index activation information.
On top of the achieved feasible complexity, the latter novel property also motivates massive \ac{IM} schemes as a promising enabler of interference-free \ac{ISAC}, by conducting information encoding only in the index domain whereas the resource-occupying signals can be utilized for sensing.

The effectiveness of the proposed framework is validated by simulations under the piloted \ac{GQSM} variant of \ac{IM}, which shows a reduced complexity where the full simulation results of unprescedented massive \ac{MIMO} \ac{IM} sizes (upto $96\times96$) was feasibly obtained with even a personal computer.
In addition, several modifications to the transmitter pilot constellation design and the detection algorithm are also provided, which further enhance the performance of the proposed detector.

{\color{black}
In all, our contributions can be summarized as follows:
\begin{itemize}
\item A novel detector for piloted \ac{IM} schemes is designed under the \ac{GaBP} framework, which enjoys a decoding complexity free of the combinatorics-dependent order. 
\item In order to enable the proposed detector, a novel \ac{UVD} formulation of \ac{IM} schemes and tailored \ac{MP} rules are presented, with the derivation of closed-form index activation \acp{PMF} shown to match the empirical distributions exactly.
\item Two enhancement modules for the proposed detector are presented, respectively based on successive \ac{IC} and a conditional probability-based denoiser, in addition to a new mechanism to optimize the symbol constellation for the \ac{GQSM} transmitter in order to improve its \ac{UVD}-based detection performance.
\item Simulation results and numerical and complexity analyses of the proposed detectors of truly massive \ac{GQSM} systems are provided and compared against the \ac{SotA} detectors and fully multiplexed \ac{MIMO}, with sizes upto $96 \times 96$ \ac{MIMO}, which is unprescedented in related literature due to the infeasible detection complexity.
\end{itemize}
}

{\color{black}
\subsubsection*{\textbf{Notation}} 
Complex scalars, vectors, and matrices are denoted by lowercase, boldface lowercase, and boldface uppercase characters, respectively, i.e., $x, \mathbf{x}, \mathbf{X}$.
$\Real{\cdot}$ and $\Imag{\cdot}$ denote the real and imaginary parts of a complex variable, while the superscripts $(\cdot)\R$ and $(\cdot)\I$ respectively denote variables which are \textit{related} to the real and imaginary parts of the system, which may still be complex-valued.
Other specific notations and functions are further defined in the text as necessary.
}


\section{System Model}
\label{sec:system_model}

As mentioned above, many \ac{SotA} \ac{IM} techniques can be found in the literature \cite{Mandloi_B5G21,Mao_CST19,Ishikawa_CST18,Cheng_WC18,Sugiura_Access17,Li_WC20,Wen_TSP15,Wen_TC17,Li_TWC22,Mesleh_TVT08,Younis_Asilomar10,Wen_JSAC19,Rou_TWC22,Raafat_TWC20,Li_Network23,Sui_TC24,Ma_JSTSP21,Gopi_TWC21,Rou_Asilomar24,Shamasundar_TWC22,Ma_TVT21,ElMai_CAMSAP23,Hu_PC19,Wang_TVT20,An_TVT22,Rou_Asilomar22_QSM,Katla_Access20,Kim_WCL21,Yukiyoshi_VTC24,Mesleh_TVT14,Xiao_TC19,Ishikawa_Access21,Xiao_TVT17,Xiao_TVT19,Nguyen_TVT18,Li_TVT22,Yang_TWC16,Cheng_TC18} which address various issues of the problem such as integration with other functionalities ($e.g.$ radar and/or sensing), achievable rates, decoding complexity and, of course which resource domains to exploit and under which architecture.
{\color{black}Without loss of generality, we exemplify the derivation and analysis of the system and signal models with the \ac{GQSM} \cite{Rou_TWC22,An_TVT22} design{\color{black}\footnote{\color{black}It will be shown in Section \ref{sec:UVD_generalised} that the proposed design is applicable to any other type of \ac{IM}-based scheme, by leveraging a \ac{UVD} reformulation.}}.}

\vspace{-2ex}
\subsection{GQSM System Model and Transmitter Structure}
\label{sec:GQSM_system_model}

Consider a point-to-point{\color{black}\footnote{\color{black}The extension of the system model and the proposed techniques to a multi-user scenario requires addressing additional non-trivial challenges, most crucially the design of multi-user \ac{IM} activation codebooks and the corresponding multi-user interference-aware detector designs, and therefore will be considered in a future work.}} (\acs{P2P}) \ac{MIMO} wireless communications system where the transmitter and the receiver are equipped with {\color{black}$N_T, N_R \!>\! 1 \!\in\! \mathbb{R}$ antennas}, respectively, such that the receive signal $\mathbf{y} \in \mathbb{C}^{N_R \times 1}$ corresponding to the transmission of the information vector $\mathbf{x} \in \mathbb{C}^{N_T \times 1}$ through the flat-fading \ac{MIMO} channel $\mathbf{H} \in \mathbb{C}^{N_R \times N_T}$ is described by
\vspace{-0.75ex}
\begin{equation}
\label{eq:received_signal}
\mathbf{y} = \mathbf{H} \mathbf{x} + \mathbf{w} \in \mathbb{C}^{N_R \times 1},
\vspace{-0.5ex}
\end{equation}
where $\mathbf{w} \in \mathbb{C}^{N_R \times 1}$ is a complex-valued \ac{AWGN} vector with element-wise variance $N_0$, such that $w_n \sim \mathcal{CN}(0, N_0)$.

In a \ac{GQSM}
transmitter, the \ac{IQ} components of $P$ transmit symbols $\mathbf{s}=\{s\cp_1,\cdots\!,s\cp_{P}\}$, taken from a complex-valued constellation $\mathcal{S}\cp$ of size $M \triangleq |\mathcal{S}\cp|$, are sparsely mapped into the respective transmit vector components $\mathbf{x}\R \in \mathbb{C}^{N_T \times 1}$ and $\mathbf{x}\I \in \mathbb{C}^{N_T \times 1}$ of the form
\vspace{-1.25ex}
\begin{subequations}
\label{eq:GQSM_transmit_signal}
\begin{eqnarray}
&\mathbf{x}\R \!\!\!\!&= [0,\cdots\!, 0, \hspace{-3ex}\overbrace{s\R_1}^{k\R_1\text{-th position}}\hspace{-3ex}, 0, \cdots\!, 0, \hspace{-3ex}\overbrace{s\R_p}^{k\R_p\text{-th position}}\hspace{-3ex}, 0,\cdots\!,0, \hspace{-3ex}\overbrace{s\R_P}^{k\R_P\text{-th position}}\hspace{-3ex}, 0]\trans, ~~\\[-0.5ex]
&\mathbf{x}\I \!\!\!\!&= [0, \hspace{-2.8ex}\underbrace{s\I_1}_{k\I_1\text{-th position}}\hspace{-2.85ex}, 0, \cdots\!, 0, \hspace{-3ex}\underbrace{s\I_p}_{k\I_p\text{-th position}}\hspace{-2.8ex}, 0,\cdots\hspace{-0.25ex},0, \hspace{-2.8ex}\underbrace{s\I_P}_{k\I_P\text{-th position}}\hspace{-3.2ex}, 0, \cdots\!, 0]\trans,~~
\end{eqnarray}

\vspace{-1ex}
\noindent yielding the transmit signal vector
\vspace{-1ex}
\begin{equation}
\mathbf{x} = \mathbf{x}\R + j\mathbf{x}\I \in \mathbb{C}^{N_T \times 1}.
\label{eq:GQSM_transmit_signal_combined}
\vspace{-1ex}
\end{equation}
\end{subequations}

The positions of the \ac{IQ} vector components in equation \eqref{eq:GQSM_transmit_signal} are described by two independent index vectors $\mathbf{k}\R \!\triangleq\! [k\R_1, \cdots, k\R_p, \cdots, k\R_P]\trans$ and $\mathbf{k}\I \!\triangleq\! [k\I_1, \cdots, k\I_p, \cdots, k\I_P]\trans$ from the set of possible index vectors $\mathcal{K}$ of size $Q \triangleq |\mathcal{K}|$.
Specifically, $Q \triangleq 2^{\lfloor\log_2\binom{N_T}{P}\rfloor}$, which implies that only binary-encodable subsets of the $\binom{N_T}{P}$ possible activation pattern combinations are utilized, such that the size of the \ac{GQSM} \ac{IQ} \emph{activation vectors} codebook $\mathcal{A}$, is given by $|\mathcal{A}| = Q$.

%
%


\vspace{1ex}
\subsubsection*{Codebook Example}

For the sake of clarity, consider a small system with $N_T = 5, P = 3$ with $\binom{5}{3} \!=\! 10$ activation vectors: \vspace{-0.5ex}
\begin{equation}
\label{eq:full_sp_codebook}
\quad
\left\{ \!\!
\begin{bmatrix}
s\x_1 \\[-0.2ex] s\x_2 \\[-0.2ex] s\x_3 \\[-0.2ex] 0 \\[-0.2ex] 0 
\end{bmatrix}\!\!,\!\!
\begin{bmatrix}
s\x_1 \\[-0.2ex] s\x_2 \\[-0.2ex] 0 \\[-0.2ex] s\x_3 \\[-0.2ex] 0 
\end{bmatrix}\!\!,\!\!
\begin{bmatrix}
s\x_1 \\[-0.2ex] s\x_2 \\[-0.2ex] 0 \\[-0.2ex] 0 \\[-0.2ex] s\x_3
\end{bmatrix}\!\!,\!\!
\begin{bmatrix}
s\x_1 \\[-0.2ex] 0 \\[-0.2ex] s\x_2 \\[-0.2ex] s\x_3 \\[-0.2ex] 0 
\end{bmatrix}\!\!,\!\!
\begin{bmatrix}
s\x_1 \\[-0.2ex] 0 \\[-0.2ex] s\x_2 \\[-0.2ex] 0 \\[-0.2ex] s\x_3
\end{bmatrix}\!\!,\!\!
\begin{bmatrix}
s\x_1 \\[-0.2ex] 0 \\[-0.2ex] 0 \\[-0.2ex] s\x_2 \\[-0.2ex] s\x_3 
\end{bmatrix}\!\!,\!\!
\begin{bmatrix}
0 \\[-0.2ex] s\x_1 \\[-0.2ex] s\x_2 \\[-0.2ex] s\x_3\\[-0.2ex] 0 
\end{bmatrix}\!\!,\!\!
\begin{bmatrix}
0 \\[-0.2ex] s\x_1 \\[-0.2ex] s\x_2 \\[-0.2ex] 0 \\[-0.2ex] s\x_3
\end{bmatrix}\!\!,\!\!
\begin{bmatrix}
0 \\[-0.2ex] s\x_1 \\[-0.2ex] 0 \\[-0.2ex] s\x_2 \\[-0.2ex] s\x_3
\end{bmatrix}\!\!,\!\!
\begin{bmatrix}
0 \\[-0.2ex] 0 \\[-0.2ex] s\x_1 \\[-0.2ex] s\x_2 \\[-0.2ex] s\x_3 
\end{bmatrix}\!\!
\right\}\!\!, \nonumber
\end{equation}
where the non-zero elements $s\x_p$ denote either $s\R_p$ or $s\I_p$, since the signal codebooks for both \ac{IQ} parts are identical.

\newpage

A valid codebook for such a system is a subset $\mathcal{A}$ of the above, containing $Q \triangleq 2^{\lfloor \log_2\binom{5}{3}\rfloor} = 8$ activation vectors, \textit{e.g.,}
\begin{equation}
\label{eq:trun_sp_codebook}
\mathcal{A}\! \triangleq\! \{ \mathbf{a}_q \}_{q = 1}^{Q = 8} 
\!=\!
\left\{ \!\!
\begin{bmatrix}
s\x_1 \\[-0.2ex] s\x_2 \\[-0.2ex] s\x_3 \\[-0.2ex] 0 \\[-0.2ex] 0 
\end{bmatrix}\!\!,\!\!
\begin{bmatrix}
s\x_1 \\[-0.2ex] s\x_2 \\[-0.2ex] 0 \\[-0.2ex] 0 \\[-0.2ex] s\x_3
\end{bmatrix}\!\!,\!\!
\begin{bmatrix}
s\x_1 \\[-0.2ex] 0 \\[-0.2ex] s\x_2 \\[-0.2ex] s\x_3 \\[-0.2ex] 0 
\end{bmatrix}\!\!,\!\!
\begin{bmatrix}
s\x_1 \\[-0.2ex] 0 \\[-0.2ex] s\x_2 \\[-0.2ex] 0 \\[-0.2ex] s\x_3
\end{bmatrix}\!\!,\!\!
\begin{bmatrix}
s\x_1 \\[-0.2ex] 0 \\[-0.2ex] 0 \\[-0.2ex] s\x_2 \\[-0.2ex] s\x_3 
\end{bmatrix}\!\!,\!\!
\begin{bmatrix}
0 \\[-0.2ex] s\x_1 \\[-0.2ex] s\x_2 \\[-0.2ex] s\x_3\\[-0.2ex] 0 
\end{bmatrix}\!\!,\!\!
\begin{bmatrix}
0 \\[-0.2ex] s\x_1 \\[-0.2ex] s\x_2 \\[-0.2ex] 0 \\[-0.2ex] s\x_3
\end{bmatrix}\!\!,\!\!
\begin{bmatrix}
0 \\[-0.2ex] 0 \\[-0.2ex] s\x_1 \\[-0.2ex] s\x_2 \\[-0.2ex] s\x_3 
\end{bmatrix}\!\!
\right\}\!\!,
\end{equation}
which can be also represented by a corresponding set of \emph{index vectors}, each of which contains the indices of the non-zero elements in the corresponding activation vector $\mathbf{a}_q$, that is
\begin{align}
\mathcal{K} &\triangleq \{ \mathbf{k}_q \}_{q = 1}^{Q = 8}  =\!
\left\{ \!\!
\begin{bmatrix}
1 \\ 2 \\ 3
\end{bmatrix}\!\!,\!\!
\begin{bmatrix}
1 \\ 2 \\ 5
\end{bmatrix}\!\!,\!\!
\begin{bmatrix}
1 \\ 3 \\ 4
\end{bmatrix}\!\!,\!\!
\begin{bmatrix}
1 \\ 3 \\ 5
\end{bmatrix}\!\!,\!\!
\begin{bmatrix}
1 \\ 4 \\ 5
\end{bmatrix}\!\!,\!\!
\begin{bmatrix}
2 \\ 3 \\ 4
\end{bmatrix}\!\!,\!\!
\begin{bmatrix}
2 \\ 3 \\ 5
\end{bmatrix}\!\!,\!\!
\begin{bmatrix}
3 \\ 4 \\ 5
\end{bmatrix}\!\!
\right\}\!.
\label{eq:codebook_K}
\end{align}

It can be understood that obtaining a given codebook $\mathcal{A}$ out of the larger set of $\binom{N_T}{P}$ possible activation vectors is a subject of optimization, and has been addressed, for instance, under a channel diversity criterion in \cite{Rou_TWC22}, and under a decoding complexity criterion via \ac{QC} in \cite{Ishikawa_Access21}.

\vspace{-2ex}
\subsection{Reformulated System Model}

In view of the formulation of \ac{GQSM} transmit signals described in equation \eqref{eq:GQSM_transmit_signal}, the received signal model in equation \eqref{eq:received_signal} can be rewritten in the \ac{IQ}-decoupled form as
\begin{equation}
\label{eq:IQ_received_signal}
\underbrace{
\begin{bmatrix}
\Re\{\mathbf{y}\} \\
\Im\{\mathbf{y}\} \\
\end{bmatrix}}_{\triangleq \;\! \bm{y}} =
\underbrace{\begin{bmatrix}
\Re\{\mathbf{H}\} & \!\!\!\!-\Im\{\mathbf{H}\}\\
\Im\{\mathbf{H}\} & \Re\{\mathbf{H}\} \\
\end{bmatrix}}_{\triangleq \;\! \bm{H}} 
\underbrace{
\begin{bmatrix}
\mathbf{x}\R \\
\mathbf{x}\I \; \\
\end{bmatrix}}_{\triangleq \;\! \bm{x}}
+
\underbrace{
\begin{bmatrix}
\Re\{\mathbf{w}\} \\
\Im\{\mathbf{w}\} \\
\end{bmatrix}}_{\triangleq \;\! \bm{w}},
\end{equation}
where $\bm{y} \in \mathbb{R}^{2N_R\times 1}$, $\bm{x} \in \mathbb{R}^{2N_T \times 1}$, and $\bm{w} \in \mathbb{R}^{2N_R \times 1}$ denote the \ac{IQ}-decoupled counterparts of $\mathbf{y}$, $\mathbf{x}$, and $\mathbf{w}$, respectively, and the \ac{IQ}-decoupled channel matrix $\bm{H} \in \mathbb{R}^{2N_R \times 2N_T}$ is further decomposed into two submatrices as
\vspace{-1ex}
\begin{equation}
\label{eq:IQ_channel}
\bm{H} \!=\! 
\Bigg[
\overbrace{
\begin{bmatrix}
\Re\{\mathbf{H}\} \\ \Im\{\mathbf{H}\}
\end{bmatrix}}^{\triangleq \bm{H}\R}
\overbrace{
\begin{bmatrix}
-\Im\{\mathbf{H}\} \\ \;\;\;\Re\{\mathbf{H}\}
\end{bmatrix}}^{\triangleq \bm{H}\I}
\Bigg] 
\!=\! \Big[\bm{H}\R ~\bm{H}\I\Big],
\end{equation}
with $\bm{H}\R \in \mathbb{R}^{2N_R \times N_T}$ and  $\bm{H}\I \in \mathbb{R}^{2N_R \times N_T}$ denoting the effective channel components for $\mathbf{x}\R$ and $\mathbf{x}\I$, respectively.

The real domain \ac{IQ}-decoupled form $\bm{y} = \bm{H}\bm{x} + \bm{w}$ presented as equation \eqref{eq:IQ_received_signal} is the basis of most \ac{QSM}/{GQSM} detection algorithms, which generally seek to solve the recovery problem
\begin{equation}
\hat{\bm{x}}_{\mathrm{ML}} = \underset{\bm{x} \in \mathcal{X}}{\mathrm{argmin}} \;\! || \bm{y} - \bm{H}\bm{x} ||_2^2,
\label{eq:ML_estimator}
\end{equation}
where $\mathcal{X}$ is the discrete domain of the effective signal $\bm{x}$, of cardinality $R \triangleq |\mathcal{X}| = \big(2^{\lfloor\log_2\!\binom{N_T}{P}\rfloor}\big)^{\!2}\!\! \cdot M^P$.

The problem described by equation \eqref{eq:ML_estimator} is NP-hard due to the discrete solution domain $\mathcal{X}$, which prevents classical convex optimization approaches \cite{Boyd_2004}, requiring instead a combinatorial search over a high-dimensional space for optimal solution.
To circumvent this challenge, \ac{SotA} low-complexity solutions have been proposed by leveraging various characteristics of the \ac{GQSM} signal.
For example, the inherent sparsity of $\bm{x}$ can be exploited by \ac{CS}-based methods \cite{Xiao_TVT17,Xiao_TVT19}, and the discrete codebook structure can be utilized to design search-based methods such as sphere decoders and pruned-search decoders \cite{Nguyen_TVT18, Wang_TVT20, Li_TVT22}.
Finally, a vector-valued \ac{MP} method was recently proposed in \cite{Rou_Asilomar22_QSM}, where the two variables $\mathbf{x}\R$ and $\mathbf{x}\I$ are concurrently estimated by leveraging the \ac{i.i.d.} bivariate property of the decoupled vectors in equation \eqref{eq:IQ_channel}.
The bivariate vector-valued \ac{MP} method of \cite{Rou_Asilomar22_QSM} was shown to reduce the search space from $R = (2^{\lfloor\log_2\!\binom{N_T}{P}\rfloor})^2 \!\cdot\! M^P$ to $2\sqrt{R} = 2\cdot (2^{\lfloor\log_2\!\binom{N_T}{P}\rfloor}) \!\cdot\! (\!\sqrt{M\;\!})^P.$

Building on the above, we propose in the sequel a novel low-complexity decoder, which leverages a new \ac{UVD}-based \ac{GQSM} system model integrated into the \ac{GaBP} framework with purpose-derived index activation probability distributions, to achieve a remarkably low decoding complexity order that is \ul{entirely independent} of the combinatorial factor $\binom{N_T}{P}$.

\section{Proposed UVD-based System Reformulation}
\label{sec:proposed_reformulation}

In this section, the  proposed \acf{UVD} of the \ac{GQSM} system model and the derivation of the novel index activation probability distributions arising from this decomposition are presented.
As shall be clarified, the resultant index-wise distributions enable the design of an \ac{MP} algorithm operating on a combinatorics-free signal domain.

\vspace{-1.5ex}
\subsection{Unit Vector Decomposition (UVD) of GQSM Scheme}
\label{sec:UVD}

First, notice that each \ac{IQ} symbol component $s\R_p$ and $s\I_p$ occupies a single position in the \ac{GQSM} transmit vector $\mathbf{x}$ described in equation \eqref{eq:GQSM_transmit_signal}, which can therefore be rewritten as a superposition of the symbol components multiplied by \textit{elementary activation vectors} $\mathbf{e}_t$, yielding \vspace{-0.5ex}
\begin{equation}
\mathbf{x} = \underbrace{\Big(\sum_{p=1}^{P} s\R_p \mathbf{e}_{k\R_p}\!\Big)}_{\triangleq \;\mathbf{x}\R}  + j\underbrace{\Big(\sum_{p=1}^{P} s\I_p \mathbf{e}_{k\I_p}\!\Big)}_{\triangleq \;\mathbf{x}\I} \in \mathbb{C}^{N_T \times 1},
\label{eq:GQSM_txvec}
\end{equation}
where $\mathbf{e}_t \in \mathbb\{0,1\}^{N_T}$, with $t \in \mathcal{T} \triangleq \{1,\cdots\!,N_T\}$, is the $t$-th column of an $N_T \times N_T$ identity matrix $\mathbf{I}_{N_T}$, which defines the set $\mathcal{E}$ of orthonormal unit vectors on $\{0,1\}^{N_T}$ as
\begin{equation}
\mathcal{E} \triangleq \{\mathbf{e}_t\}_{t=1}^{N_T} =
\left\{ \!
\begin{bmatrix}
1 \\ 0 \\[-0.5ex] \vdots \\[0.5ex] 0 
\end{bmatrix}\!\!,
\begin{bmatrix}
0 \\ 1 \\[-0.5ex] \vdots \\[0.5ex] 0 
\end{bmatrix}\!\!,
\cdots,
\begin{bmatrix}
0 \\ 0 \\[-0.5ex] \vdots \\[0.5ex] 1 
\end{bmatrix}\!
\right\}\!.
\label{eq:set_UV}
\end{equation}
\vspace{-2ex}
\begin{figure}[H]
\centering
\includegraphics[width=0.6\columnwidth]{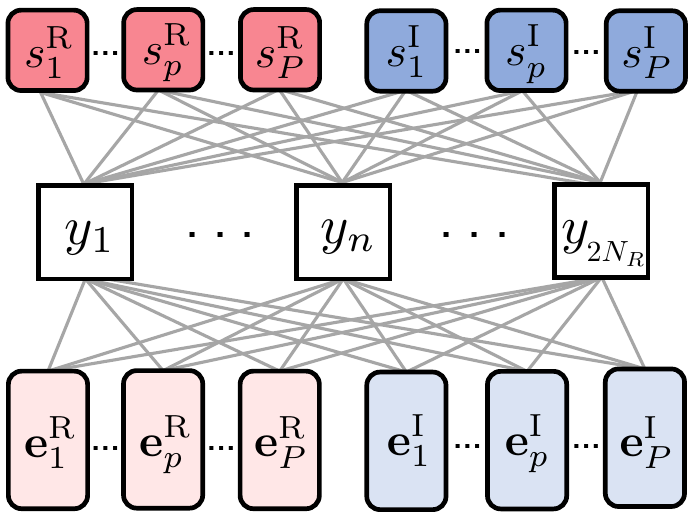}%
\caption{Multivariate bilinear factor graph with $2P$ symbol variables and $2P$ unit vector variables, as per equation \eqref{eq:reformulated_IQ_received_signal}.}
\label{fig:fg_3}
\end{figure}

Then, by reformulating the transmit vector $\mathbf{x}$, the \ac{IQ}-decoupled received signal vector of equation \eqref{eq:IQ_received_signal} becomes
\begin{equation}
\label{eq:reformulated_IQ_received_signal}
\bm{y} = \bm{H}
\begin{bmatrix}
\sum_{p=1}^{P} s\R_p \mathbf{e}_{k\R_p} \\
\!\sum_{p=1}^{P} s\I_p \mathbf{e}_{k\I_p} \\
\end{bmatrix}
+
\bm{w} \in \mathbb{R}^{2N_R \times 1},
\end{equation}
which better highlights the construct of the \ac{GQSM} signals with inherently independent spatial and digital encoding, as illustrated by the bilinear factor graph shown in Fig. \ref{fig:fg_3}.

In other words, the estimation of the full effective signal $\bm{x}$ has been transformed into the \ul{joint} estimation problem of $2P$ activation vectors and $P$ transmit symbols, which also enables independent estimation techniques when some of the decoupled variables are known, \textit{i.e.,} via pilot symbols.

\subsection{\ac{UVD} Model for Coded and Space-time Coded IM}
\label{sec:UVD_generalised}

The \ac{UVD} model described above works for all \ac{SM} schemes whose activation vectors are directly given by the unit vectors in $\mathcal{E}$, $e.g.$, \ac{GSM} and \ac{GQSM}.
There exists, however, enhanced \ac{SM}-based schemes \cite{Yang_TWC16, Cheng_TC18} whose transmit symbols are not multiplied by elementary activation vectors in the space of $\{0,1\}^{N_T}$, but by optimized precoded \textit{dispersion vectors} drawn from a complex space $\mathbb{C}^{N_T}$.

Let then $\mathcal{D}\R \triangleq \{\mathbf{d}\R_q\}_{q=1}^{Q} \in \mathbb{C}^{N_T \times 1}$ and $\mathcal{D}\I \triangleq \{\mathbf{d}\I_q\}_{q=1}^{Q} \in \mathbb{C}^{N_T \times 1}$ be the sets from which the precoded dispersion vectors of a generic coded \ac{SM} are drawn.
Then, the corresponding transmit signal vector is given by
\begin{equation}
\mathbf{x} = \Big(\sum_{p=1}^{P} s\R_p \mathbf{d}\R_{k\R_p}\!\Big) + j\Big(\sum_{p=1}^{P} s\I_p \mathbf{d}\I_{k\I_p}\!\Big) \in \mathbb{C}^{N_T \times 1}.
\label{eq:tx_signal_precoded}
\end{equation}

Next, we leverage the trivial identities
\begin{equation}
~~~~\mathbf{d}\R_q \triangleq 
\overbrace{\begin{bmatrix}
\mathbf{d}\R_1, \cdots\!, \mathbf{d}\R_Q
\end{bmatrix}}^{\triangleq \;\! \mathbf{\Xi}\R \;\! \in \;\! \mathbb{C}^{N_T \!\times\! Q}} \!\cdot\,
\mathbf{e}_q 
~~\text{and}~~ 
\mathbf{d}\I_q \triangleq 
\overbrace{\begin{bmatrix}
\mathbf{d}\I_1, \cdots\!, \mathbf{d}\I_Q
\end{bmatrix}}^{\triangleq \;\! \mathbf{\Xi}\I \;\! \in \;\! \mathbb{C}^{N_T \!\times\! Q}}
\!\cdot\,
\mathbf{e}_q,
\label{eq:dispersion_vector_selection}
\end{equation}
with $\mathbf{e}_q \in \{0,1\}^{Q \times 1}$, to rewrite equation \eqref{eq:tx_signal_precoded} as
\begin{equation}
\mathbf{x} =
\mathbf{\Xi}\R
\Big(\!\sum_{p=1}^{P} s\R_p \mathbf{e}_{k\R_p}\!\Big) + j\mathbf{\Xi}\I \Big( \!\sum_{p=1}^{P} s\I_p \mathbf{e}_{k\I_p}\!\Big) \in \mathbb{C}^{N_T \times 1},
\label{eq:generalized_tx}
\end{equation}
where $\mathbf{\Xi}\R \in \mathbb{C}^{N_T \times Q}$ and $\mathbf{\Xi}\I \in \mathbb{C}^{N_T \times Q}$ are the dispersion vector codebook matrices constructed by concatenating the $Q$ dispersion vectors in $\mathcal{D}\R$ and $\mathcal{D}\I$ respectively\footnote{It can be seen that the formulation for the \ac{GQSM} system described by equations \eqref{eq:GQSM_txvec} and \eqref{eq:reformulated_IQ_received_signal} is the simplified case of this generalized \ac{UVD} form, as the matrices $\mathbf{\Xi}\R$ and $\mathbf{\Xi}\I$ reduce to the identity matrix in the \ac{GQSM} case.}

Then, the \ac{IQ}-decoupled form of the received signal corresponding to the transmit vector in equation \eqref{eq:generalized_tx} becomes
\begin{eqnarray}
\bm{y} \hspace{-4ex} && = \!\bm{H}\!
\overbrace{\begin{bmatrix}
\Re\{\mathbf{\Xi}\R\} & \!\!\!\!\!-\Im\{\mathbf{\Xi}\I\}\\
\Im\{\mathbf{\Xi}\R\} & \!\Re\{\mathbf{\Xi}\I\} \\
\end{bmatrix}}^{\triangleq \;\! \bm{\varXi} \in \mathbb{R}^{2N_T \times 2Q}} \!
\begin{bmatrix}
\sum_{p=1}^{P} s\R_p \mathbf{e}_{k\R_p} \\
\!\sum_{p=1}^{P} s\I_p \mathbf{e}_{k\I_p} \\
\end{bmatrix}
\!+\!
\bm{w},\nonumber\\
\label{eq:generalized_reformulated_IQ_received_signal_simple}
&& = \!\tilde{\bm{H}}\begin{bmatrix}
\sum_{p=1}^{P} s\R_p \mathbf{e}_{k\R_p} \\
\!\sum_{p=1}^{P} s\I_p \mathbf{e}_{k\I_p} \\
\end{bmatrix}\!+\!
\bm{w} \in \mathbb{R}^{2N_R \times 1},
\vspace{-1ex}
\end{eqnarray}
where the effective channel $\tilde{\bm{H}} \triangleq \bm{H} \bm{\varXi} \in \mathbb{R}^{2N_R \times 2Q}$ is constructed from the dispersion codebook $\bm{\varXi}$ and the \ac{CSI} $\bm{H}$, both known at the receiver. 

The structural equivalence between equations  \eqref{eq:generalized_reformulated_IQ_received_signal_simple} and \eqref{eq:reformulated_IQ_received_signal} is clear, from which it is evident that the decoding method to be described in Section \ref{sec:proposed_decoder} can be applied to both cases.
Following similar steps, it can be also shown that \ac{IM} schemes incorporating \ac{STC}, such as those proposed in \cite{Rou_TWC22,Wang_TVT20}, in which dispersion vectors are generalized into dispersion matrices of size ${N_T \times T}$ with $T$ consecutive symbol periods, can also be represented, via a vectorization of the system model, by the same \ac{UVD} formulation shown above, once again enabling the proposed technique of Section \ref{sec:proposed_decoder} to be applied.

To elaborate, given the sets of dispersion \textit{matrices} $\mathcal{D}\R \triangleq \{\mathbf{D}\R_q\}_{q=1}^{Q} \in \mathbb{C}^{N_T \times T}$ and $\mathcal{D}\I \triangleq \{\mathbf{D}\I_q\}_{q=1}^{Q} \in \mathbb{C}^{N_T \times T}$, the transmit signal of a \ac{STC}-based \ac{IM} scheme becomes

\quad\\[-4ex]
\begin{equation}
\mathbf{X} = \bigg(\sum_{p=1}^{P} s\R_p \mathbf{D}\R_{k\R_p}\!\bigg) + j\bigg(\sum_{p=1}^{P} s\I_p \mathbf{D}\I_{k\I_p}\!\bigg) \in \mathbb{C}^{N_T \times T},
\label{eq:tx_signal_STC}
\vspace{-1ex}
\end{equation}
whose vectorized form is trivially given as \vspace{-0.5ex}
\begin{align}
\mathrm{vec}(\mathbf{X}) \!=\!\! \bigg(\!\sum_{p=1}^{P} \!s\R_p \mathrm{vec}(\mathbf{D}\R_{k\R_p})\!\bigg) \!\!+\! j\bigg(\!\sum_{p=1}^{P} \!s\I_p \mathrm{vec}(\mathbf{D}\I_{k\I_p})\!\bigg) \!\!\in\! \mathbb{C}^{TN_T \times 1}. \nonumber \\[-3.5ex]
\label{eq:tx_signal_STC_vec}
\end{align}

\vspace{-0.5ex}
The \ac{UVD} representation of equation \eqref{eq:tx_signal_STC_vec} is therefore similar to that of  equation \eqref{eq:generalized_tx}, except for the increased dimensionality to $TN_T$, and the dictionary matrices of vectorized dispersion matrices $\mathbf{\Xi}\R \triangleq [\mathrm{vec}(\mathbf{D}\R_1),\cdots\!,\mathrm{vec}(\mathbf{D}\R_Q)] \in \mathbb{C}^{TN_T \times Q}$ and $\mathbf{\Xi}\I \triangleq [\mathrm{vec}(\mathbf{D}\I_1),\cdots\!,\mathrm{vec}(\mathbf{D}\I_Q)] \in \mathbb{C}^{TN_T \times Q}$, which yields the vectorized received signal model \vspace{-0.5ex}
\begin{align}
\bm{y} & \triangleq \mathrm{vec}(\bm{Y}) = \mathrm{vec}(\bm{H}\!\bm{X} \!+\! \bm{W}),\nonumber \\
& = \!(\mathbf{I}_T \!\otimes\! \bm{H})\!
\begin{bmatrix}
\Re\{\mathbf{\Xi}\R\} & \!\!\!\!\!-\Im\{\mathbf{\Xi}\I\}\\
\Im\{\mathbf{\Xi}\R\} & \!\Re\{\mathbf{\Xi}\I\} \\
\end{bmatrix} \!\!
\begin{bmatrix}
\sum_{p=1}^{P} s\R_p \mathbf{e}_{k\R_p} \\
\!\sum_{p=1}^{P} s\I_p \mathbf{e}_{k\I_p} \\
\end{bmatrix}
\!+\!
\mathrm{vec}({\bm{W}}), \nonumber\\
\label{eq:generalized_reformulated_IQ_received_signal_vec}
& = \!\tilde{\bm{H}}\begin{bmatrix}
\sum_{p=1}^{P} s\R_p \mathbf{e}_{k\R_p} \\
\!\sum_{p=1}^{P} s\I_p \mathbf{e}_{k\I_p} \\
\end{bmatrix}\!+\!\tilde{\bm{w}}\in \mathbb{R}^{2TN_R \times 1},
\end{align}
where $\mathbf{I}_T$ is the $T \times T$ identity matrix, the symbol $\otimes$ denotes Kronecker product, and $\tilde{\bm{H}}$ and $\tilde{\bm{w}}$ are the augmented effective channel matrix and \ac{AWGN} vector, respectively.

From the equivalence among equations \eqref{eq:generalized_reformulated_IQ_received_signal_vec}, \eqref{eq:generalized_reformulated_IQ_received_signal_simple} and \eqref{eq:reformulated_IQ_received_signal}, it follows once again that the \ac{UVD} model introduced in \eqref{sec:UVD} for the \ac{GQSM} scheme generalizes to other \ac{IM} methods {in other domains and higher dimensions}.
{\color{black} Therefore, we emphasize again that all of the following derivations, analyses, and proposed methods are applicable to all general types of \ac{IM}.
However, for the sake of simplicity, the remainder of the article will continue to use the \ac{GQSM} scheme as an exemplary case.}
\vspace{-4ex}

\vspace{-1ex}
\subsection{Prior Distribution of Unit Vector Variables}
\label{sec:prior_distribution}
\vspace{-0.5ex}

Let us now proceed to analyze the prior distributions of the two sets of variables in the \ac{UVD} model, namely the \ac{IQ} symbol components $s\R_p$ and $s\I_p$, and the elementary activation vectors $\mathbf{e}_{k\R_p}$ and $\mathbf{e}_{k\I_p}$, respectively, in order to enable their estimation via Bayesian inference methods.

Since \ac{QSM} schemes typically employ complex symmetric constellations\footnotemark ~such as $M$-\ac{QAM}, $s\R_p$ and $s\I_p$ and belong to the corresponding real-valued effective constellation ${\mathcal{S}} \triangleq \Re\{\mathcal{S}\cp\} \!=\! \Im\{\mathcal{S}\cp\}$, which is  typically a \ac{PAM} constellation of cardinality $|{\mathcal{S}}| \triangleq \sqrt{M}$.

Succinctly, we therefore have
\begin{equation}
s\R_p, s\I_p  \sim \mathbb{P}_{\mathsf{s}}(s_p) \triangleq \sum_{s \in \mathcal{S}} \frac{1}{\sqrt{M}} \!\cdot \delta(s - s_p).
\label{eq:PMF_sp}  \vspace{-0.5ex}
\end{equation}

On the other hand, the \ac{PMF} of $\mathbf{e}_{k\R_p}$ and $\mathbf{e}_{k\I_p}$ is not so trivial, and is given by \vspace{-0.5ex}
\begin{equation}
\mathbf{e}_{k\R_p}, \mathbf{e}_{k\I_p} \!\sim\!
\mathbb{P}_{\mathsf{e_{k_p}\!}}(\mathbf{e}_{k_p}) \triangleq \! \sum_{\mathbf{e}_t \in \mathcal{E}} \!\mathbb{P}_{\mathsf{e_{k_p}\!}}(\mathbf{e}_{k_p} \!\!= \mathbf{e}_t) \cdot \delta(\mathbf{e}_{k_p} \!- \mathbf{e}_t),
\label{eq:PMF_ekp} \vspace{-0.5ex}
\end{equation}
which can be alternatively described solely in terms of the corresponding activation indices $k\R_p$ and $k\I_p$, such that equation \eqref{eq:PMF_ekp} can equivalently be written as \vspace{-0.5ex}
\begin{equation}
k\R_p, k\I_p  \sim \mathbb{P}_{\mathsf{k_p\!}}(k_p) \triangleq \sum_{t \in \mathcal{T}} \mathbb{P}_{\mathsf{{k_p}\!}}({k_p} = t) \cdot \delta(k_p - t).
\label{eq:PMF_kp} \vspace{-0.5ex}
\end{equation}

\begin{figure}[b!]
\vspace{-2.5ex}
\centering
\includegraphics[width=0.975\columnwidth]{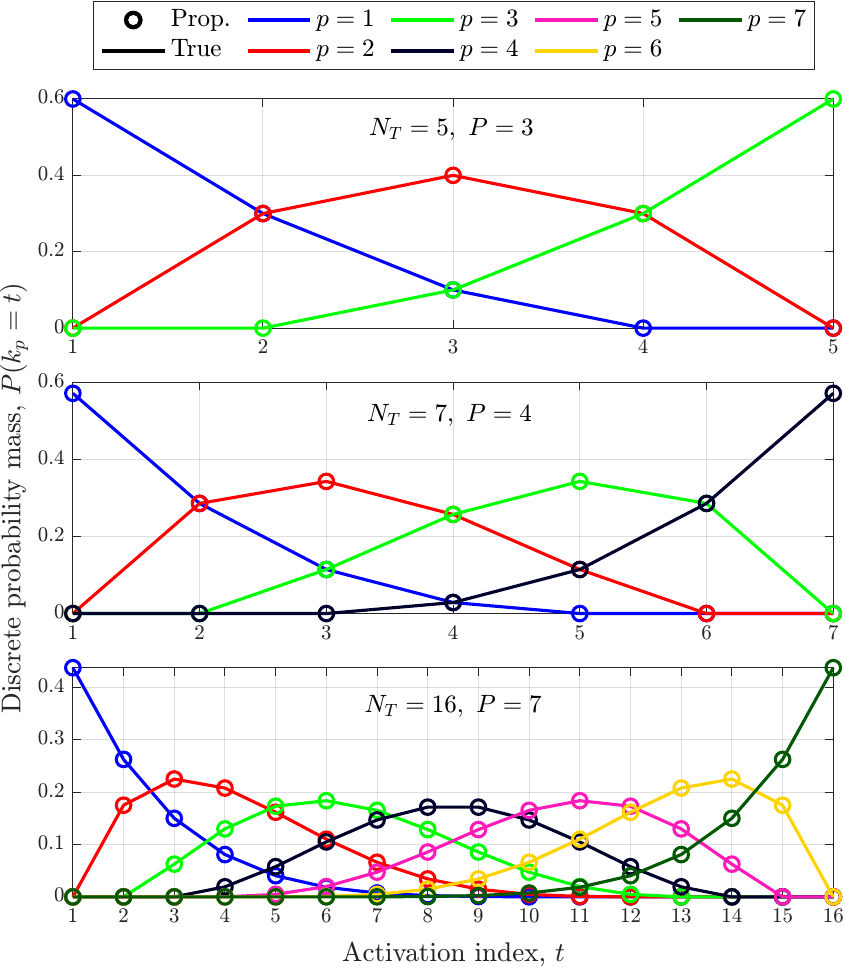}
\caption{Comparison of discrete polynomial-based \ac{PMF} and empirically-obtained distribution of the activation indices of the index vectors in $\mathcal{K}$.}
\label{fig:PMF_model_dist}
\end{figure}

The challenge in equation \eqref{eq:PMF_kp} lies in the probabilities $\mathbb{P}_{\mathsf{k_p\!}}(k_p)$, which far from being uniform and independent, is strongly correlated and differs significantly for each $p$, as a consequence of the combinatorial selection which yields finite and ordered index lists.
To illustrate, consider the example shown in equation \eqref{eq:codebook_K} for $N_T = 5$ and $P = 3$, from which it can be seen that the first element $k_1$ has a higher probability of taking on the value of $1$, then $2$ and $3$, and can never take on the values $4$ or $5$.
Similar restrictions apply for the other index vector elements $k_2$ and $k_3$.

Fortunately, the analytical \acp{PMF} corresponding to the index vector elements $k_p,\forall p$, can be derived using tools of numerical and combinatorial analysis \cite{LinzBook2019, JohnRiordanBook2002, Sablonniere_JCAM93}.
In particular, consider the following bounded variations of Bernstein polynomials \cite{Sablonniere_JCAM93} on the discrete variable $t \in \mathcal{T} \triangleq \{1,\cdots\!,N_T\}$ given by

$~$
\vspace{-3.5ex}
\begin{eqnarray}
\label{eq:novel_poly}
{R}^{P}_p(t; N_T) \hspace{-4ex} && = \\
&& \hspace{-5ex} =
\begin{cases}
\! \dfrac{(t\!-\!1) (N_T \!-\! t)!}{(t\!-\!p)! \big((N_T \!-\! t) \!-\! (P \!-\! p)\big)!} &  \!\!\text{if } t \text{ satisfies eq. \eqref{eq:poly_nonzero_condition},} \\[1ex]
\;\!0 & \!\!\text{otherwise,}
\end{cases} \nonumber 
\end{eqnarray} \vspace{-2ex}

\noindent where \vspace{-1ex}
\begin{equation}
\label{eq:poly_nonzero_condition}
(p - 1) < t < (N_T - P - p + 1). \vspace{-1ex}
\end{equation}

For each value $p$, the polynomial ${R}^{P}_p(t; N_T)$ gives the number of the mutually exclusive index $t$, out of a list of length $P$, confined to the interval $\mathcal{T}$, such that the desired distributions are obtained by merely normalizing such polynomials, yielding \vspace{-1ex}
\begin{equation}
\label{eq:PMF_model}
\mathbb{P}_{\mathsf{{k_p}\!}}({k_p} = t) = \dfrac{{R}^P_p(t;N_T)}{\textstyle{\sum}_{t \in \mathcal{T}} \,\! {R}^P_p(t;N_T)} \triangleq \mathbb{P}_{\mathsf{{e_{k_p}}\!}}(\mathbf{e}_{k_p} \!= \mathbf{e}_t).
\end{equation}

\footnotetext{If that is not the case, and $\mathcal{S}\cp$ is irregular, the formulation is trivially generalized by considering distinct PAM constellations for $s\R_p$ and $s\I_p$.}

Some examples of the \ac{PMF} given by equation \eqref{eq:PMF_model} are compared to the true empirical distributions in Figure \ref{fig:PMF_model_dist}, for various $N_T$ and $P$, which confirm that the expression is exact.

For future convenience, the probability masses of the \ac{PMF} $
\mathbb{P}_{\mathsf{e_{k_p}\!}}(\mathbf{e}_{k_p})$ is represented by the vector $\mathbf{r}_p$ defined as \vspace{-0.5ex}
\begin{equation}
\mathbf{r}_p 
\triangleq [\mathbb{P}_{\mathsf{{e_{k_p}}\!}}(\mathbf{e}_{k_p} \!\triangleq\! \mathbf{e}_1), \cdots\!, \mathbb{P}_{\mathsf{{e_{k_p}}\!}}(\mathbf{e}_{k_p} \!= \mathbf{e}_{N_T})]\trans \in \mathbb{R}^{N_T \times 1}.\!
\label{eq:vec_PMF}
\end{equation}

\vspace{-1.5ex}
\section{\color{black}Proposed Low-Complexity Detector \\ for Pilotted Massive Index Modulation Schemes}
\label{sec:proposed_decoder}

With possession of the analytical prior \acp{PMF} of the indices $k_p$ given in equations \eqref{eq:PMF_sp} and \eqref{eq:PMF_model}, respectively, we next design a \ac{GaBP}-based \ac{MP} algorithm for their estimation. 
In particular, we will derive the \ac{MP} rules for a piloted \ac{GQSM}, \textit{i.e.,} where $s\R_p, s\I_p ~\forall p$ are known, which can be utilized to enable secondary functionalities\footnote{We highlight that, in principle, a bilinear \ac{MP} decoder for this system could also be developed \cite{Parker_TSP14,Rou_Asilomar22_JCAS, Rou_TWC24, Ito_TC23}, such that some of the pilots could be replaced by information-carrying symbols to increase the rate of communications, albeit at the expense of some performance degradation. Such extension is out of scope of the current article and will be pursued in a follow-up work.} such as sensing \cite{Gaudio_TWC20, Rou_Asilomar22_JCAS, Ranasinghe_ICASSP24, Ranasinghe_Arxiv24}, as well as communication via the detection of activation patterns. 

First, the soft-replicas for the activation vector variables $\mathbf{e}_{k\R_p}$ and $\mathbf{e}_{k\I_p}$ with $p \in \{1, \cdots, P\}$ are defined for the $n$-th factor node ($i.e., n$-th pilot) with $n \in \{1, \cdots, 2N_R\}$, as $\srvR$ and $\srvI$, respectively, {\color{black}as illustrated in the factor graph of Figure \ref{fig:fg_3}.}
The corresponding error covariance matrices of the activation vector soft-replicas are defined as
\begin{subequations}
\begin{align}
\mathbf{\Gamma}\R_{p:n} &\triangleq \Exp_{\mathsf{e_{k_p}}\!}[(\mathbf{e}_{k_p} \!\!- \srvR)(\mathbf{e}_{k_p} \!\!- \srvR)\herm] \in \mathbb{R}^{N_T \times N_T}, \\
&= \textstyle{\sum}_{\mathbf{e}_t \in \mathcal{E}} \big( \mathbb{P}_{\mathsf{e_{k_p}\!\!}}(\mathbf{e}_{k_p} \!\!= \mathbf{e}_t) \cdot (\mathbf{e}_t \!\!- \srvR)(\mathbf{e}_t \!\!- \srvR)\herm \big), \nonumber \\
\mathbf{\Gamma}\I_{p:n} &\triangleq \Exp_{\mathsf{e_{k_p}}\!}[(\mathbf{e}_{k_p} \!\!- \srvI)(\mathbf{e}_{k_p} \!\!- \srvI)\herm] \in \mathbb{R}^{N_T \times N_T}, \\
&= \textstyle{\sum}_{\mathbf{e}_t \in \mathcal{E}} \big( \mathbb{P}_{\mathsf{e_{k_p}\!\!}}(\mathbf{e}_{k_p} \!\!= \mathbf{e}_t) \cdot (\mathbf{e}_t \!- \srvI)(\mathbf{e}_t \!- \srvI)\herm \big), \nonumber \vspace{-2ex}
\end{align}
\label{eq:error_covariance}%
where the expectation is taken over the corresponding unique \acp{PMF} of equation \eqref{eq:PMF_model}, for each $p \in \{1,\cdots\!,P\}$.
\end{subequations}

\vspace{1ex}

\textit{\textbf{Remark:} It can be seen from equation \eqref{eq:error_covariance} and the system analysis in Section \ref{sec:system_model}, that the derivations corresponding to the two \ac{IQ} components are identical.
Therefore, for the sake of brevity, we hereafter derive expressions only for the real components, with the understanding that the same results apply to the imaginary components, only with the terms $(\,\cdot\,)\R$ replaced by $(\,\cdot\,)\I$.}

\vspace{0.5ex}

\begin{figure*}[b]
    \setcounter{equation}{28}
    \centering
    \vspace{-1ex}
    \hrulefill
    \vspace{-1ex}
    \begin{equation}
    \bar{y}\R_{p:n} \triangleq \overbrace{(\mathbf{h}_n\R)\transs \! s_{p}\R \;\! \srvR}^{\text{True symbol}} + \overbrace{(\mathbf{h}_n\R)\transs \!\!\textstyle\sum\limits_{p' \neq p}^{P} \!\! s_{p'}\R (\avRp \!\!- \srvRp) + (\mathbf{h}_n\I)\transs \!\!\textstyle\sum\limits_{p'=1}^{P} \!\! s_{p'}\I (\avIp \!\!- \srvIp) + w_n}^{\text{Scalar Gaussian approximation (\ac{SGA})}},
    \label{eq:soft-ic-SGA}
    \end{equation}
    \end{figure*}

    \setcounter{equation}{25}
Computed naively, the covariances in equation \eqref{eq:error_covariance} requires a summation of $N_T$ outer products of dimensions $N_T \times N_T$, with complexity $\mathcal{O}(N_T^3)$.
Thanks to the \ac{UVD} and the sparsity of the unit vectors in $\mathcal{E}$, however, equation \eqref{eq:error_covariance} can be put into the following closed-form of associated complexity $\mathcal{O}(N_T^2)$,
\begin{equation}
\mathbf{\Gamma}\R_{p:n} = \diag{\mathbf{r}_p} + \srvR \srvR\trans - (\mathbf{E}\R_{p:n} + (\mathbf{E}\R_{p:n})\transs),
\label{eq:closed-form_covmat}
\end{equation} 
%
%
where the $N_T \times N_T$ square matrix $\mathbf{E}\I_{p:n}$ is given by
\begin{equation}
\mathbf{E}\R_{p:n} \triangleq \mathbf{r}_p\trans \otimes \srvR = [{r}_{p}{_{(1)}}\!\cdot\!\srvR, \,\cdots, {r}_{p}{_{(N_T)}}\!\cdot\!\srvR],
\label{eq:E_matrix}%
\end{equation}
%
with ${r}_p{_{(t)}}$ denoting the $t$-th element of the vector of probability mass coefficients of $\mathbf{r}_p$ from equation \eqref{eq:vec_PMF}.

In hand of the soft-replica vectors and the error covariance matrices, the $2N_R$ factor nodes corresponding to the received symbols in $\bm{y}$, each perform soft-\ac{IC} on the received symbols $y_n$ for the activation vectors as
\begin{equation}
\bar{y}\R_{p:n}\! = \!y_n \!-\! (\mathbf{h}_n\R)\transs \! \sum_{p' \neq p}^{P} (s_{p'}\R \srvRp) - (\mathbf{h}_n\I)\transs \! \sum_{p' = 1}^{P} (s_{p'}\I \srvIp),
\label{eq:soft-ic}
\end{equation}
%
%
where $(\mathbf{h}_n\R)\transs\! \!\in\! \mathbb{R}^{1 \times N_T}$ and $(\mathbf{h}_n\I)\transs \!\!\in\! \mathbb{R}^{1 \times N_T}$ respectively denote the $n$-th rows of the channel components $\bm{H}\R \in \mathbb{R}^{2N_T \times N_T}$ and $\bm{H}\I \in \mathbb{R}^{2N_T \times N_T}$ which are defined in equation \eqref{eq:IQ_channel}.

As can be seen in equation \eqref{eq:soft-ic-SGA} at the bottom of the page, the soft-\ac{IC} symbol is comprised of the true symbol part and a term related to the soft-replica error plus \ac{AWGN}, such that the latter term can be approximated by a Gaussian scalar via the \ac{CLT}, which yields the conditional \acp{PDF} of the soft-\ac{IC} symbols conditional to a given activation vector as
\setcounter{equation}{29}
\begin{equation}
\mathbb{P}(\bar{y}_{p:n}\R|\mathbf{e}_{k\R_p}) \propto \exp\!\bigg( \!-\! \frac{|\bar{y}_{p:n}\R - s_{p}\R (\mathbf{h}_n\R)\transs \!\mathbf{e}_{k\R_p}|^2}{\nu\R_{p:n}} \bigg),
\label{eq:conditional_pdf}
\end{equation}
%
%
where the conditional variance $\nu\R_{p:n}$ is obtained via
\begin{align}
\nu\R_{p:n} & = \Exp_{\mathsf{e_{k_p}}\!\!}\!\left[ |\bar{y}_{p:n}\R - s_{p}\R (\mathbf{h}_n\R)\transs\! \mathbf{e}_{k\R_p}|^2 \right] \nonumber \\
& = \nu_{n} - (\mathbf{h}_n\R)\transs\! \big( |s_{p}\R|^2 \cdot \mathbf{\Gamma}_{p:n}\R \big) (\mathbf{h}_n\R)^{*} + \tfrac{N_0}{2},
\label{eq:conditional_variances}
\end{align}    
%
%
with the total variance $\nu_{n}$ prior to the soft-\ac{IC}
\begin{align}
\nu_{n} \!\triangleq\! (\mathbf{h}_n\R)\transs \!\big(\!\!\textstyle\sum\limits_{p = 1}^{P}\!\!|s_{p}\R|^2\!  \cdot \!\mathbf{\Gamma}_{p:n}\R \big)(\mathbf{h}_n\R)^{*} \!\!+\!(\mathbf{h}_n\I)\transs\!\big(\!\!\textstyle\sum\limits_{p = 1}^{P}\!\!|s_{p}\I|^2 \!\cdot\! \mathbf{\Gamma}_{p:n}\I \big) (\mathbf{h}_n\I)^{*}\! \nonumber \\[-2ex]
\end{align}

\vspace{-1ex}
The conditional \acp{PDF} of equation \eqref{eq:conditional_pdf} are combined at the variable nodes to obtain the extrinsic belief $b_{p:n}\R$, \textit{i.e.,}
\vspace{-1ex}
\begin{align}
\mathbb{P}(b_{p:n}\R|\mathbf{e}_{k\R_p}) & = \!\!\prod_{n' \neq n}^{2N_R} \!\!\mathbb{P}(\bar{y}_{p:n'}\R|\mathbf{e}_{k\R_p}) \label{eq:extrisinc_pdf} \\[-1ex] 
& ~~~~~\propto \mathrm{exp}\big((\bm{\eta}\R_{p:n})\transs \mathbf{e}_{k\R_p} - \tfrac{1}{2}(\mathbf{e}_{k\R_p})\transs \mathbf{\Lambda}\R_{p:n} \mathbf{e}_{k\R_p}\big),  \nonumber
\end{align}
where self-\ac{IC} is included for the computation of the extrinsic belief for the $n$-th factor node, as can be seen by the exclusion of the $n$-th conditional \ac{PDF} from the variable node.

The resulting unscaled multivariate Gaussian \ac{PDF} of the extrinsic beliefs are efficiently described in terms of the information vector $\bm{\eta}_{p:n}\R \in \mathbb{R}^{N_T \times 1}$, which is given by \vspace{-1ex}
\begin{equation}
\bm{\eta}_{p:n}\R = s\R_{p} \!\sum_{n' \neq n}^{2N_R} \!\frac{\bar{y}\R_{p:n'}}{\nu\R_{p:n'}}\mathbf{h}_{n'}\R,
\label{eq:information_vector} \vspace{-1ex}
\end{equation}
and in terms of the precision matrix $\mathbf{\Lambda}_{p:n}\R \!\in\! \mathbb{R}^{N_T \!\times\! N_T} \!$ given by \vspace{-1.25ex}
\begin{equation}
\mathbf{\Lambda}_{p:n}\R = |s_p\R|^2 \!\! \sum_{n' \neq n}^{2N_R} \!\frac{\mathbf{h}_{n'}\R (\mathbf{h}_{n'}\R)\transs}{\nu\R_{p:{n'}}}.
\label{eq:precision_matrix} \vspace{-1ex}
\end{equation}

In turn, the posterior Bayes-optimal soft-replicas are computed from the extrinsic beliefs via \vspace{-1ex}
\begin{align}
\srvR & = \!\dfrac{\Exp_\mathsf{e_{k_p}\!}[\mathbf{e}_{k_p} \!\!\cdot \mathbb{P}(b_{p:n}\R|\mathbf{e}_{k_p})]}{\Exp_\mathsf{e_{k_p}\!}[\mathbb{P}(b_{p:n}\R|\mathbf{e}_{k_p})]} \in \mathbb{R}^{N_T \times 1} \label{eq:posterior_soft-replica} \\
& = \frac{\sum_{\mathbf{e}_t \in \mathcal{E}} \! \big( \mathbf{e}_t \cdot \mathbb{P}_{\mathsf{e_{k_p}\!\!}}(\mathbf{e}_{k_p} \!\!= \mathbf{e}_t) \cdot \mathbb{P}(b_{p:n}\R|\mathbf{e}_{k_p}) \big)}{\sum_{\mathbf{e}'_t \in \mathcal{E}} \! \big( \mathbb{P}_{\mathsf{e'_{k_p}\!\!}}(\mathbf{e}'_{k_p} \!\!= \mathbf{e}'_t) \cdot \mathbb{P}(b_{p:n}\R|\mathbf{e}'_{k_p}) \big)}, \nonumber  \vspace{-1.5ex}
\end{align}  
%

Similarly to equation \eqref{eq:error_covariance}, the expectation of $\mathbf{e}_{k_p}$ over its domain $\mathcal{E}$ can be efficiently described in a closed-form expression with reduced computational complexity, namely \vspace{-1ex}
\begin{equation}
\srvR \!=\! \frac{\mathbf{r}_{p} \!\odot\! {\mathbf{z}\R_{p:n}}}{\mathbf{r}_{p}\trans \cdot {\mathbf{z}\R_{p:n}}} \in \mathbb{R}^{N_T \times 1},
\label{eq:posterior_soft-replica_closedform}
\vspace{-1ex}
\end{equation}
where the vector of unnormalized belief mass $\mathbf{z}\R_{p:n}$ is given by \vspace{-3ex}
\begin{align}
{\mathbf{z}}\R_{p:n} & = {\mathrm{exp}}_{\odot} \!\big( \bm{\eta}\R_{p:n} - \tfrac{1}{2} \diag{\mathbf{\Lambda}\R_{p:n}}\! \big)  \in \mathbb{R}^{N_T \times 1},    \label{eq:posterior_soft-replica_closedform_es} \\
& =[e^{{\eta}\R_{p:n}{_{(1)}} - \frac{1}{2}{\Lambda}\R_{p:n}{_{(1,1)}}}, \cdots,
e^{{\eta}\R_{p:n}{_{(N_T)}} - \frac{1}{2}{\Lambda}\R_{p:n}{_{(N_T,N_T)}}}]\trans\!, \nonumber 
\end{align}
%
\vspace{-3.5ex}

\noindent with ${\mathrm{exp}}_{\odot}\!(\cdot)$ denoting the element-wise exponentiation of a vector, ${\eta}\R_{p:n}{_{(t)}}$ denoting the elements at the $t$-th position of the vector $\bm{\eta}\R_{p:n}$, and ${\Lambda}\R_{p:n}{_{(t,t)}}$ denoting the elements at the $(t,t)$-th postion of the matrix $\mathbf{\Lambda}\R_{p:n}$. 

Notice that equation \eqref{eq:posterior_soft-replica_closedform_es} clarifies that only the diagonal elements of the precision matrix are required for the computation of the soft-replica vectors.
Therefore, instead of computing the entire precision matrix as in equation \eqref{eq:precision_matrix}, the vector of diagonal elements can be directly obtained to significantly improve the computational complexity, given by \vspace{-1ex}
\begin{equation}
\diag{\;\!\mathbf{\Lambda}_{p:n}\R} = |s_p\R|^2 \! \sum_{n' \neq n}^{2N_R} \!\frac{|\mathbf{h}_{n'}\R|^2_{\!\odot}}{\nu\R_{p:{n'}}}
\label{eq:diagonal_precision_matrix} \vspace{-1ex}
\end{equation}
where $|\cdot|_{\!\odot}$ denotes the element-wise absolute value operator.

Equations \eqref{eq:soft-ic} to \eqref{eq:diagonal_precision_matrix} describe the steps of one \ac{MP} iteration to estimate the $2P$ activation vectors of the piloted \ac{GQSM} with \ac{UVD}, which yields the refined posterior soft-replica vectors and the corresponding error covariance matrices.

At the end of a $\tau$-th \ac{MP} iteration, the soft-replica vectors and error covariance matrices are damped \cite{Som_ITW10} in order to prevent early convergence to a local optima \cite{Su_TSP15}, thus  \vspace{-0.5ex}
\begin{equation}
\srvR^{[\tau + 1]} \longleftarrow \rho \cdot \srvR^{[\tau]} + (1-\rho) \cdot \srvR^{[\tau + 1]},\vspace{-0.5ex} 
\label{eq:damping_update}
\end{equation}   
where $\rho \in [0,1]$ is the damping factor, and the superscript $(\cdot)^{[\tau]}$ denotes the $\tau$-th iterate of the variable.

The last of the \ac{MP} iterations, denoted by the index $\tau_\mathrm{conv}$, can be determined by well-known convergence criteria, such as the maximum number of iterations, \textit{i.e.,} while $\tau \leq \tau_{\mathrm{max}}$, or a given convergence threshold of the soft-replicas \textit{i.e.,} while $\big|\srvR^{[\tau + 1]} - \srvR^{[\tau]}\big| > \varepsilon_\mathrm{Th}$.
After the last \ac{MP} iteration, a belief consensus is taken over the $2N_R$ factor nodes via \vspace{-1ex}
\begin{align}
\mathbb{P}(b_{p}\R|\mathbf{e}_{k\R_p}) & = \!\!\prod_{n = 1}^{2N_R} \!\mathbb{P}(\bar{y}_{p:n}\R|\mathbf{e}_{k\R_p}) \label{eq:consensus_pdf} \\[-1ex]
& ~~~~~\propto \mathrm{exp}\big((\bm{\eta}\R_{p})\transs \mathbf{e}_{k\R_p} - \tfrac{1}{2}(\mathbf{e}_{k\R_p})\transs \mathbf{\Lambda}\R_{p} \mathbf{e}_{k\R_p}\big),  \nonumber
\end{align}      
%
which is equivalent to equation \eqref{eq:extrisinc_pdf} without the self-interference cancellation operation.

Similarly, the consensus information vectors and precision matrix diagonals are respectively given by \vspace{-0.75ex}
\begin{equation}
\bm{\eta}_{p}\R = s\R_{p} \!\sum_{n = 1}^{2N_R} \frac{\bar{y}\R_{p:n}}{\nu\R_{p:n}}\mathbf{h}_{n}\R, \vspace{-0.75ex}
\label{eq:consensus_information_vector}
\end{equation}
\begin{equation}
\diag{\;\!\!\mathbf{\Lambda}_{p}\R} \!=\! |s_p\R|^2 \! \sum_{n = 1}^{2N_R} \!\frac{|\mathbf{h}_{n}\R|^2_{\!\odot}}{\nu\R_{p:{n}}}, \vspace{-0.25ex}
\label{eq:consensus_diagonal_precision_matrix}
\end{equation}

Finally, a hard-decision on the $2P$ activation vectors is made by evaluating the consensus \acp{PDF} for all $N_T$ valid states of the activation vectors, \textit{i.e.,} \vspace{-1ex}
\begin{equation}
\tilde{\mathbf{{e}}}_{k\R_p}  = \underset{\mathbf{e}_t \in \mathcal{E}}{\mathrm{argmax}} ~\mathbb{P}(b_{p}\R|\mathbf{e}_{t}), \vspace{-0.75ex}
\label{eq:final_decision}
\end{equation}
which is equivalent to computing the consensus soft-replicas \vspace{-0.75ex}
\begin{equation}
\hat{\mathbf{{e}}}_{k\R_p} \!=\! \frac{\mathbf{r}_{p} \!\odot\! {\mathbf{z}\R_{p}}}{\mathbf{r}_{p}\trans \cdot {\mathbf{z}\R_{p}}}, \vspace{-1ex}
\end{equation}%
where \vspace{-0.25ex}
\begin{equation}
{\mathbf{z}}\R_{p} \!=\! {\mathrm{exp}}_{\odot} \!\big( \bm{\eta}\R_{p} \!-\! \tfrac{1}{2} \diag{\mathbf{\Lambda}\R_{p}}\! \big), \vspace{-0.5ex}
\label{eq:consensus_posterior_soft-replica_closedform_es}%
\end{equation}
and determining the index of the maximum value of the consensus soft-replica, \textit{i.e.,} \vspace{-0.5ex}
\begin{equation}
k\R_p = \underset{t \in \mathcal{T}}{\mathrm{argmax}} (\hat{\mathrm{e}}_{k\R_p(t)}), \vspace{-0.5ex}
\label{eq:equivalent_hard_decision}
\end{equation}
where $\hat{\mathrm{e}}_{k\R_p(t)}$ denote the $t$-th element of $\hat{\mathbf{{e}}}_{k\R_p}$.

$~$ \vspace{-2.5ex}

\begin{algorithm}[H]
\hrulefill
\begin{algorithmic}[1]
\vspace{-1ex}
\Statex \hspace{-3ex} {\bf{Inputs:}} Received signal $\bm{y}$, 
effective channels $\bm{H}\R$ and $\bm{H}\I$, 
\Statex \hspace{4.65ex} pilot symbols $s\R_p, s\I_p \;\forall p$, and noise variance $N_0$.
\Statex \hspace{-3ex} {\bf{Outputs:}} Estimated activation vectors $\tilde{\mathbf{{e}}}_{k\R_p}$ and $\tilde{\mathbf{{e}}}_{k\I_p} \;\forall p$.
\vspace{-1.5ex}
\Statex \hspace{-4ex}\hrulefill
\Statex \hspace{-3ex} \textbf{Initialization:} $\forall n$ and $\forall p$,
\State Initialize the soft-replicas $\srvR$, $\srvI$, following the prior \ac{PMF} of equation \eqref{eq:vec_PMF};
\State Compute $\mathbf{\Gamma}\R_{p:n},\mathbf{\Gamma}\I_{p:n}$ via equation \eqref{eq:closed-form_covmat}; \vspace{0.5ex}
\Statex \hspace{-3ex} \textbf{MP iterations until convergence}, $\;\forall n$ and $\forall p$,
\State Perform soft-\ac{IC} to obtain $\bar{y}\R_{p:n}$, $\bar{y}\I_{p:n}$ via equation \eqref{eq:soft-ic};
\State Compute $\nu\R_{p:n}, \nu\I_{p:n}$ via equation \eqref{eq:conditional_variances};
\State Compute $\bm{\eta}_{p:n}\R, \bm{\eta}_{p:n}\I$ via equation \eqref{eq:information_vector};
\State Compute $\mathrm{diag}(\mathbf{\Lambda}_{p:n}\R), \mathrm{diag}(\mathbf{\Lambda}_{p:n}\I)$ via equation \eqref{eq:diagonal_precision_matrix};
\State Compute $\srvR$, $\srvI$ via equation \eqref{eq:posterior_soft-replica_closedform};
\State Compute $\mathbf{\Gamma}_{p:n}\R, \mathbf{\Gamma}_{p:n}\I$ via equation \eqref{eq:closed-form_covmat};
\State Update $\srvR$, $\srvI$, $\mathbf{\Gamma}_{p:n}\R, \mathbf{\Gamma}_{p:n}\I$ with damping via eq. \eqref{eq:damping_update};
\Statex \hspace{-3ex} \textbf{end for} \vspace{0.5ex}
\Statex \hspace{-3ex} \textbf{Belief Consensus:}
$\forall p$,
\State Obtain $\bm{\eta}_{p}\R,\bm{\eta}_{p}\I$ via equation \eqref{eq:consensus_information_vector};
\State Obtain $\mathrm{diag}(\mathbf{\Lambda}_{p}\R),\mathrm{diag}(\mathbf{\Lambda}_{p}\I)$ via equation \eqref{eq:consensus_diagonal_precision_matrix};
\State Obtain $\tilde{\mathbf{{e}}}_{k\R_p}$, $\tilde{\mathbf{{e}}}_{k\I_p}$ via equation \eqref{eq:final_decision};
\caption[]{\!\!: Proposed \acs{UVD}-\ac{GaBP} \ac{GQSM} Decoder}
\label{alg:UVD-GaBP}
\end{algorithmic}
\end{algorithm}
\setlength{\textfloatsep}{12pt}

\begin{figure}[H]
\centering
\includegraphics[width=1\columnwidth]{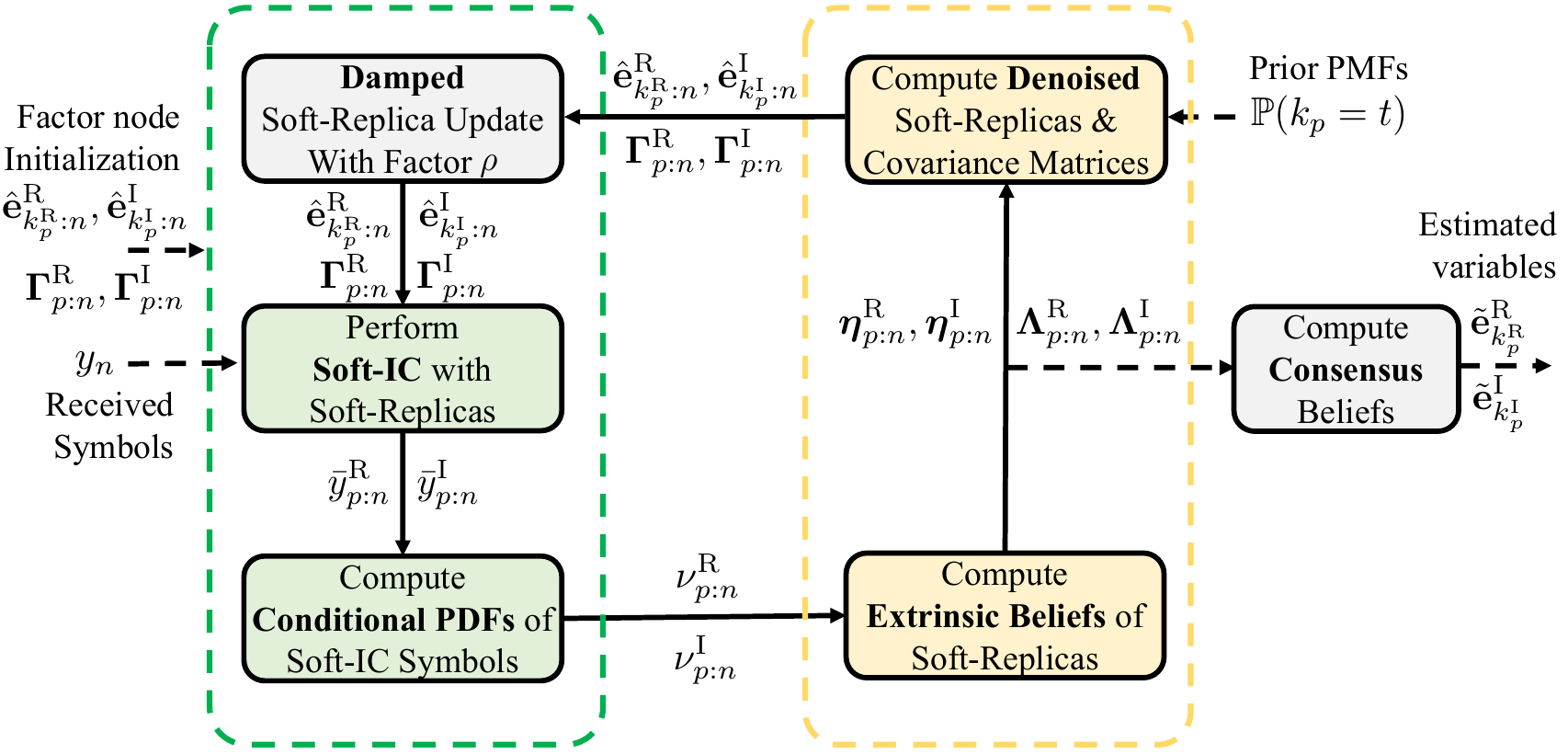}
\vspace{-3ex}
\caption{Schematic diagram of the proposed \ac{UVD}-\ac{GaBP} detector.
The operations in the factor nodes and the variable node have been illustrated in green and yellow colors, respectively.}
\label{fig:schematics}
\vspace{-0.5ex}
\end{figure}

The proposed low-complexity decoder hereby dubbed the ``\ac{UVD}-\ac{GaBP}" for piloted \ac{GQSM} schemes, described by equation \eqref{eq:error_covariance}-\eqref{eq:equivalent_hard_decision}, is summarized by Algorithm \ref{alg:UVD-GaBP} and illustrated as a schematic diagram in Figure \ref{fig:schematics}.

\vspace{-0.5ex}
\section{Improving the Proposed UVD-GaBP Decoder}
\label{sec:enhanced_decoder}

In this section, the fundamental limitations and the consequent sources of errors of the proposed decoder in Section \ref{sec:proposed_decoder} are identified, which mainly arises from the omission of the joint probability information between the unit vectors of the \ac{UVD}. 
Subsequently, enhancements to the algorithm are proposed to improve the detection performance against such errors, at the cost of an additional computational complexity yet minimal structual change.

\begin{figure}[H]
\vspace{-3ex}
\centering
\captionsetup[subfloat]{labelfont=small,textfont=small}
\subfloat[Correct convergence with uniquely estimated unit vectors.]{\includegraphics[width=0.475\columnwidth]{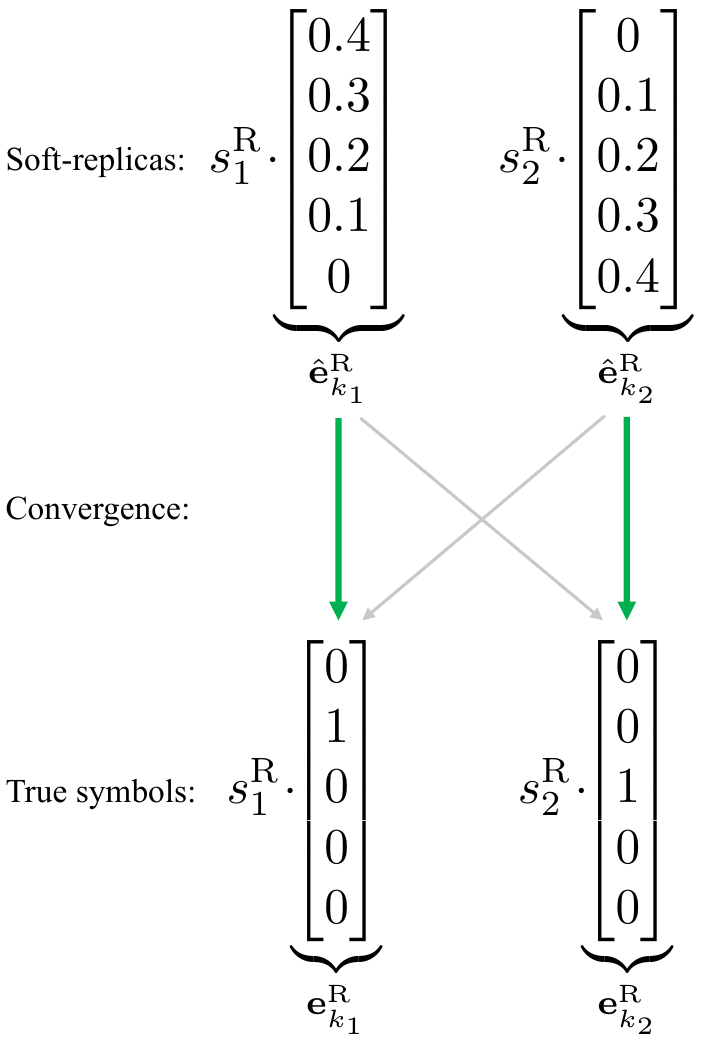}%
\label{fig:UVDconv_noerror}}
\quad
\subfloat[Erroneous convergence with duplicate unit vector estimates.]{\includegraphics[width=0.475\columnwidth]{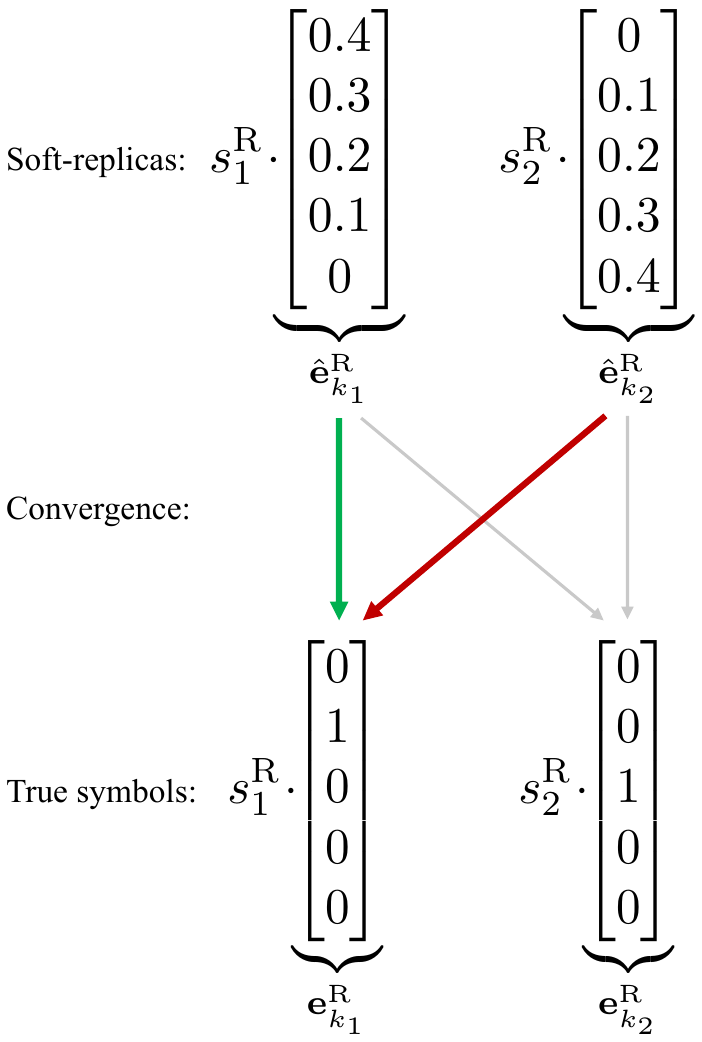}%
\label{fig:UVDconv_error}}
\vspace{-1ex}
\caption{Two exemplary cases of unit vector variable convergence, for a \ac{GQSM} system with $N_T = 5$, $P = 2$.}
\label{fig:UVD_error}
\vspace{-2ex}
\end{figure}

Consider two soft-replicas\footnote{Without loss of generality the dependencies on the observation node denoted by the subscript $(\,\cdot\,)_{:n}$ are omitted here, and only the real part variables $(\,\cdot\,)\R$ are described, as the same analysis applies to the imaginary part variables $(\,\cdot\,)\I$.} $\hat{\mathbf{e}}\R_{k_{p}}$ and $\hat{\mathbf{e}}\R_{k_{p'}}$ of two unique unit vectors $\mathbf{e}\R_{k_p}$ and $\mathbf{e}\R_{k_{p'}}$ with $p \neq p'$, and the respective transmit symbols $s\R_p$ and $s\R_{p'}$.
Correct estimation results in the convergence of the effective symbol $s\R_p \!\cdot\! \hat{\mathbf{e}}\R_{k_{p}}$ to the true value of $s\R_p \!\cdot\! {\mathbf{e}}\R_{k_{p}}$, and $s\R_{p'} \!\cdot\! \hat{\mathbf{e}}\R_{k_{p'}}$ to $s\R_{p'} \!\cdot\! {\mathbf{e}}\R_{k_{p'}}$, as illustrated in Figure \ref{fig:UVDconv_noerror}.

On the other hand, an erroneous estimation may arise when a soft-replica converges to the wrong unit vector, as illustrated in Figure \ref{fig:UVDconv_error}, leading to a \textit{duplicate} estimate.
Due to the combinatoric selection of the indices in a \ac{GQSM} index vector and of general \ac{IM} schemes as described in Section \ref{sec:GQSM_system_model}, no duplicate indices may exist within a single index vector.

In light of the above, two sources of such erroneous duplicate convergence are identified: \textit{a)} high transmit symbol similarity, and \textit{b)} lack of joint/dependent distribution information between the unit vector variables.
In order to mitigate the error from the above identified sources, three modifications to the \ac{GQSM} transmitter and the \ac{UVD}-\ac{GaBP} decoder are proposed in the following.

\vspace{-1ex}

\subsection{\ac{IQ}-Orthogonal Constellation Design}
\label{sec:pilot_constellation_design}

Let us assume that multiple transmit symbols have high similarity, \textit{i.e.,} $|s\R_p - s\R_{p'}|^2 \!\approx\! 0$.
Such behavior is illustrated by observing Fig. \ref{fig:UVDconv_error} and considering $s\R_p \approx s\R_{p'}$, where the initialized effective symbols $s\R_p \cdot \hat{\mathbf{e}}\R_{k_{p}}$ and $s\R_{p'} \cdot \hat{\mathbf{e}}\R_{k_{p'}}$ are consequently very similar in value.
Therefore, the effective symbols appearing in the core \ac{MP} steps of the \ac{UVD}-\ac{GaBP} decoder in equations \eqref{eq:soft-ic}-\eqref{eq:precision_matrix}, may converge to either of the true symbols $s\R_p \cdot {\mathbf{e}}\R_{k_{p}}$ or $s\R_{p'} \cdot {\mathbf{e}}\R_{k_{p'}}$ and potentially produce an erroneous duplicate estimate.

The first proposed approach to mitigate the duplicate convergence error is to modify the transmit symbol constellation, such that a sufficient difference in numerical value are ensured for the symbols in the respective real and imaginary domains.

\begin{figure}[H]
\vspace{-3ex}
\centering
\captionsetup[subfloat]{labelfont=small,textfont=small}
\subfloat[Symmetric \\ 4\ac{QAM} constellation.]{\includegraphics[width=0.31\columnwidth]{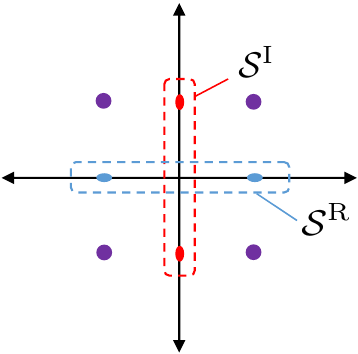}%
\label{fig:normalQAM}}
\hfil
\subfloat[Optimally rotated 4\ac{QAM} constellation.]{\includegraphics[width=0.31\columnwidth]{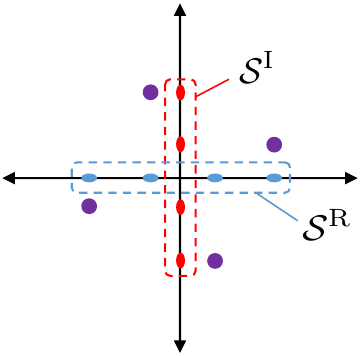}%
\label{fig:rotatedQAM}}
\hfil
\subfloat[Optimal 4\ac{QAM} within the 16\ac{QAM}.]{\includegraphics[width=0.31\columnwidth]{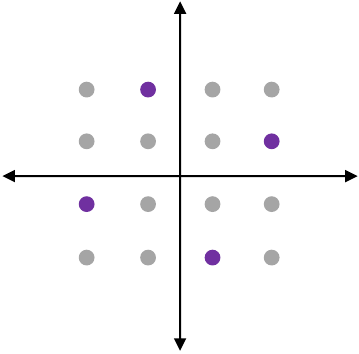}%
\label{fig:selectionQAM}}
\vspace{-0.5ex}
\caption{Illustrations of the described complex constellations, where the symbols of the constellation are denoted in purple.}
\label{fig:constell_example}
\vspace{-2ex}
\end{figure}

To that end, consider the \ac{IQ}-decoupled constellations $\mathcal{S}\R \triangleq \Re\{\mathcal{S}^\mathrm{c}\}$ and $\mathcal{S}\I \triangleq \Im\{\mathcal{S}^\mathrm{c}\}$, which is equivalent to two \ac{PAM} constellations respectively on the real and imaginary axes, as illustrated in Figure \ref{fig:normalQAM}.
For a symmetric $M$-\ac{QAM} constellation $\mathcal{S}^\mathrm{c}$, the \ac{PAM} constellations have a smaller cardinality than $M$, as multiple symbols in $\mathcal{S}^\mathrm{c}$ may convey the same amplitude in the \ac{IQ} domain, leading to duplicate symbol values.

In order to avoid such duplicate symbols in the \ac{IQ} domain, we propose a simple rotation scheme to $\mathcal{S}^\mathrm{c}$, which leads to non-duplicate \ac{PAM} symbols in both \ac{IQ} domains.
Furthermore, an optimal rotation angle $\theta^*$ is obtained to maximize the minimum Euclidean distance between the symbols of the \ac{PAM} constellation.
There exist an optimal rotation angle $\theta^*$ for each of the quadrants, but we only consider the solution within the first quadrant, \textit{i.e.,} $\theta \in (0, \frac{\pi}{2})$, which are shown in Table \ref{tab:optimal_rotationangle}.

The optimization problem is described by \vspace{-0.5ex}
\begin{equation}
\theta^* = \underset{\theta \in (0, \frac{\pi}{2})}{\mathrm{argmax}} \; D_{\mathrm{min}}\big(\Re\{e^{j\theta} \!\cdot\mathcal{S}^{\mathrm{c}}\}\big) + D_{\mathrm{min}}\big(\Im\{e^{j\theta}\!\cdot\mathcal{S}^{\mathrm{c}}\}\big), \vspace{-0.5ex}
\end{equation}
where $D_{\mathrm{min}\!}(\,\cdot\,)$ is a function on a set denoting the minimum Euclidean distance of its elements defined as \vspace{-0.5ex}
\begin{equation}
D_{\mathrm{min}\!}(\mathcal{S}) \triangleq \!\! \underset{s_{m\vphantom{'}}\!, s_{m'} \in \mathcal{S}}{\mathrm{min}} \, \sqrt{|s_{m\vphantom{'}} \!- s_{m'}|^2}, ~\forall m \neq m'. \vspace{-0.5ex}
\end{equation}

It is observed that the optimal rotation illustrated in Figure \ref{fig:rotatedQAM} is effectively equivalent to fitting a \ac{QAM} constellation from $M$ unique \ac{IQ}-orthogonal symbols from the a larger scaled \ac{QAM} constellation to maximize the minimum Euclidean distances between the symbols, as illustrated in Figure \ref{fig:selectionQAM}.

In light of the above, given that the pilot symbols are selected uniquely, the rotated constellation ensures no duplicate or \textit{similar} symbols in each of the decoupled \ac{IQ} domain.

\begin{table}[H]
\vspace{1ex}
\centering
\caption{Optimal rotation angles $\theta^*$ in radians (to the nearest 3 decimal points) for different constellation sizes $M$.}
\vspace{-0.5ex}
\begin{tabular}[t]{|c||c|c|c|c|c|c|}
\hline\hline
$M$ & 4 & 16 & 32 & 64 & 128 & 256 \\ \hline
$\theta^*$ & 0.464 & 0.245 & 0.165 & 0.124 & 0.082 & 0.062 \\ \hline\hline
\end{tabular}
\label{tab:optimal_rotationangle}
\vspace{-1ex}
\end{table}

\subsection{Conditional \ac{PDF}-based Denoiser}
\label{sec:conditional_denoiser}

In addition to the optimal rotation of the symbol constellation which is a pre-computed at the \ac{GQSM} transmitter, another approach is proposed to enhance the proposed detector in this section.
In order to avoid the duplicate estimation error of the unit vectors the dependent information of the unit vectors can be incorporated in the form of conditional \acp{PMF}.
Specifically, the prior \acp{PMF} $\mathbb{P}_{\mathsf{e_{k_p}\!\!}}(\mathbf{e}_{k_p} \!\!=\! \mathbf{e}_t)$ of the Bayes-optimal soft-replica denoiser in equation \eqref{eq:posterior_soft-replica} may be enhanced to first-order\footnote{While it is also possible to incorporate higher-order conditional \acp{PMF} by incorporating more than one conditional variable, the derivation and application of such higher-order \acp{PMF} require a significantly increased computational and space complexity, such that only the first-order conditional \ac{PMF} is considered in this work for practicality.}s conditional \acp{PMF} $\mathbb{P}_{\mathsf{e_{k_p} | \mathsf{k}_{{\mathsf{p}'}}}\!}(\mathbf{e}_{k_p} \!\!=\! \mathbf{e}_t \,|\, \mathbf{e}_{k_{p'}} = \mathbf{e}_{t'}) \equiv \mathbb{P}_{\mathsf{e_{k_p} | \mathsf{k}_{{\mathsf{p}'}}}\!}(\mathbf{e}_{k_p} \!\!=\! \mathbf{e}_t \,|\, k_{p'} = t')$, for $p \neq p'$ and $t \neq t'$.
In other words, the first-order conditional \acp{PMF} yield the activation probability of the $t$-th index of the $p$-th unit vector, given the activation information of $t'$-th index of the $p'$-th unit vector.

However, the evaluation of a conditional \ac{PMF} requires hard-decided activation indices for a given unit vector, which is not available within the \ac{MP} iterations fundementally based on soft-replicas.
Therefore within each iteration of the \ac{MP} algorithm, a greedy hard-determination of the most confident index pair ($p$, $t$) is made for each $n$-th factor node by evaluating equation \eqref{eq:extrisinc_pdf}, and finding the index pair yielding the maximal belief mass, \textit{i.e.,} by solving \vspace{-0.5ex}
\begin{equation}
\check{p}\R_{n}, \check{t}\R_{n} = \underset{p,t}{\mathrm{argmax}} \;\mathbb{P}(b_{p:n}\R|\mathbf{e}_{t}), 
\label{eq:greedy_index} \vspace{-1ex}
\end{equation} 
which is equivalent to obtaining the indices of the maximum values in the belief mass vector $\mathbf{z}\R_{p:n}$ in equation \eqref{eq:posterior_soft-replica_closedform_es}.

Under such greedy selection, the conditional \ac{PMF}-based denoiser with the greedy indices $\check{p}\R_{n}, \check{t}\R_{n}$, becomes
\begin{align}
\srvR & = \!\dfrac{\Exp_\mathsf{e_{k_p} | \mathsf{k}_{\check{\mathsf{p}}}}\![\mathbf{e}_{k_p} \!\!\cdot \mathbb{P}(b_{p:n}\R|\mathbf{e}_{k_p}) \,|\, k\R_{\check{p}} = \check{t}\R_n \,]}{\Exp_\mathsf{e_{k_p} | \mathsf{k}_{\check{\mathsf{p}}}}\![\mathbb{P}(b_{p:n}\R|\mathbf{e}_{k_p})\,|\, k\R_{\check{p}} = \check{t}\R_n \, ]} \in \mathbb{R}^{N_T \times 1},  \label{eq:conditional_posterior_soft-replica}  \\
& = \frac{\sum_{\mathbf{e}_t \in \mathcal{E}} \! \big( \mathbf{e}_t \cdot \mathbb{P}_\mathsf{e_{k_p} | \mathsf{k}_{\check{\mathsf{p}}}}\!(\mathbf{e}_{k_p} \!\!=\! \mathbf{e}_t \,|\, k_{\check{p}}\! =\! \check{t}_n) \!\cdot\! \mathbb{P}(b_{p:n}\R|\mathbf{e}_{k_p}) \big)}{\sum_{\mathbf{e}'_t \in \mathcal{E}} \! \big( \mathbb{P}_\mathsf{e_{k_p} | \mathsf{k}_{\check{\mathsf{p}}}}\!(\mathbf{e}'_{k_p} \!\!= \mathbf{e}'_t \,|\, k_{\check{p}} \!=\! \check{t}_n) \cdot \mathbb{P}(b_{p:n}\R|\mathbf{e}'_{k_p} ) \big)}, \nonumber
\end{align}  

\noindent to yield the soft-replica vectors incorporating the dependent information of the most confident unit vector variable.

The conditional \acp{PMF} does not exhibit a closed-form expression unlike equation \eqref{eq:PMF_model} for the independent \acp{PMF}, and therefore must be obtained via numerical evaluation.
However, as the conditional \acp{PMF} are fixed for a given codebook, \textit{i.e.,} system parameters $N_T$ and $P$, all values can be pre-computed.

\vspace{-2ex}
\subsection{Greedy Successive Interference Cancellation (\ac{IC})}
\label{sec:greedy_IC}
\vspace{-0.25ex}

Furthermore, a similar approach can also be performed after the convergence of the \ac{MP} iterations to find the index of the strongest value within the converged beliefs in equation \eqref{eq:consensus_pdf} \vspace{-0.75ex}
\begin{equation}
\check{p}\R, \check{t}\R = \underset{p,t}{\mathrm{argmax}} \;\mathbb{P}(b_{p}\R|\mathbf{e}_{t}), \label{eq:greedy_index_conv} \vspace{-0.75ex}
\end{equation} 
which is equivalent to obtaining the indices of the maximum values in the belief masses $\mathbf{z}\R_{p}$ in equation \eqref{eq:consensus_posterior_soft-replica_closedform_es}.

In hand of a converged index of the most confident unit vector variable, \ac{IC} can be performed on the received signals corresponding to the selected indices as \vspace{-0.75ex}
\begin{equation}
\check{y}\R_{n} \triangleq y_n - (\mathbf{h}_{n}\R)\transs(s\R_{\check{p}\R_n} \!\cdot\! \mathbf{e}_{\check{t}\R_n}), \vspace{-0.75ex}
\label{eq:greedy_sic} 
\end{equation}
and the corresponding variables are nullified from the respective soft-replica, to be repeated for $P$ successive \ac{IC} iterations.

Finally, the estimated index vectors can be obtained by concatenating and sorting the inherent activation indices from each greedy selection step without without having to evaluate equation \eqref{eq:final_decision}-\eqref{eq:equivalent_hard_decision}, \textit{i.e.,} \vspace{-0.5ex}
\begin{equation}
\tilde{\mathbf{k}}\R = \mathrm{sort}([{\check{t}\R_{[1]}}, \cdots, {\check{t}\R_{[P]}}]), \vspace{-1ex}
\label{eq:index_vector_eval}
\end{equation}  
where $\check{t}\R_{[\lambda]}$ denote the greedily selected activation index at the $\lambda$-th iteration of the successive \ac{IC} loop.

$~$ \vspace{-3ex}

\begin{algorithm}[H]
\hrulefill
\begin{algorithmic}[1]
\vspace{-1ex}
\Statex \hspace{-3ex} {\bf{Inputs:}} Received signal $\bm{y}$, 
effective channels $\bm{H}\R$ and $\bm{H}\I$, 
\Statex \hspace{4.65ex} pilot symbols $s\R_p, s\I_p \;\forall p$, and noise variance $N_0$.
\Statex \hspace{-3ex} {\bf{Outputs:}} Estimated index vectors $\tilde{\mathbf{k}}\R$ and $\tilde{\mathbf{k}}\I$.
\vspace{-1.5ex}
\Statex \hspace{-4ex}\hrulefill \vspace{-0.5ex}
\Statex \hspace{-3ex} \textbf{Initialization:} $\forall n$ and $\forall p$,
\State Initialize the soft-replicas $\srvR$, $\srvI$, following the prior \ac{PMF} of equation \eqref{eq:vec_PMF};
\State Compute $\mathbf{\Gamma}\R_{p:n},\mathbf{\Gamma}\I_{p:n}$ via equation \eqref{eq:closed-form_covmat}; \vspace{0.5ex}
\Statex \hspace{-3ex} \textbf{Successive IC iterations for} $\lambda = 1,\cdots, P$,
\Statex \hspace{-1ex} \textbf{MP iterations until convergence}, $\;\forall n$ and $\forall p$,
\State \hspace{0.25ex} Perform soft-\ac{IC} to obtain $\bar{y}\R_{p:n}$, $\bar{y}\I_{p:n}$ via equation \eqref{eq:soft-ic};
\State \hspace{0.25ex} Compute $\nu\R_{p:n}, \nu\I_{p:n}$ via equation \eqref{eq:conditional_variances};
\State \hspace{0.25ex} Compute $\bm{\eta}_{p:n}\R, \bm{\eta}_{p:n}\I$ via equation \eqref{eq:information_vector};
\State \hspace{0.25ex} Compute $\mathrm{diag}(\mathbf{\Lambda}_{p:n}\R), \mathrm{diag}(\mathbf{\Lambda}_{p:n}\I)$ via equation \eqref{eq:diagonal_precision_matrix};
\State \hspace{0.25ex} Evaluate $\check{p}\R_{n}, \check{p}\I_{n}$ and $\check{t}\R_{n}, \check{t}\I_{n}$ via equation \eqref{eq:greedy_index};
\State \hspace{0.25ex} Compute $\srvR$, $\srvI$ via equation \eqref{eq:conditional_posterior_soft-replica};
\State \hspace{0.25ex} Compute $\mathbf{\Gamma}_{p:n}\R, \mathbf{\Gamma}_{p:n}\I$ via equation \eqref{eq:error_covariance};
\State \hspace{0.25ex} Update $\srvR$, $\srvI$, $\mathbf{\Gamma}_{p:n}\R, \mathbf{\Gamma}_{p:n}\I$ via damping eq. \eqref{eq:damping_update};
\Statex \hspace{-1ex} \textbf{end for} 
\Statex \hspace{-1ex} \hspace{-1.5ex} \textbf{Belief Consensus:}
$\forall p$,
\State Obtain $\bm{\eta}_{p}\R,\bm{\eta}_{p}\I$ via equation \eqref{eq:consensus_information_vector};
\State Obtain $\mathrm{diag}(\mathbf{\Lambda}_{p}\R),\mathrm{diag}(\mathbf{\Lambda}_{p}\I)$ via equation \eqref{eq:consensus_diagonal_precision_matrix};
\State Evaluate $\check{p}\R, \check{p}\I$ and $\check{t}\R, \check{t}\I$ via equation \eqref{eq:greedy_index_conv};
\State Store the values of $\check{p}\R, \check{p}\I$ and $\check{t}\R, \check{t}\I$ at the $\lambda$-th \ac{IC} iteration.
\State Perform \ac{IC} to obtain $\check{y}\R_{n}, \check{y}\I_{n}$ via equation \eqref{eq:greedy_sic};
\Statex \hspace{-3ex} \textbf{end for} \vspace{0.2ex}
\Statex \hspace{-3ex} \textbf{Index Vector Evaluation:}
\State Obtain $\tilde{\mathbf{k}}\R,\tilde{\mathbf{k}}\I$ via equation \eqref{eq:index_vector_eval};
\caption[]{\!\!: Proposed \acs{UVD}-\ac{GaBP} \ac{GQSM} Decoder \\ $~$ ~~~~~~~~~~~~~~~ with Conditional Denoiser and Successive \ac{IC}}
\label{alg:UVD-GaBP_SIC_Cond}
\end{algorithmic}
\end{algorithm}
\setlength{\textfloatsep}{12pt}
\vspace{-3.5ex}
\newpage

\begin{figure}[H]
\centering
\includegraphics[width=1\columnwidth]{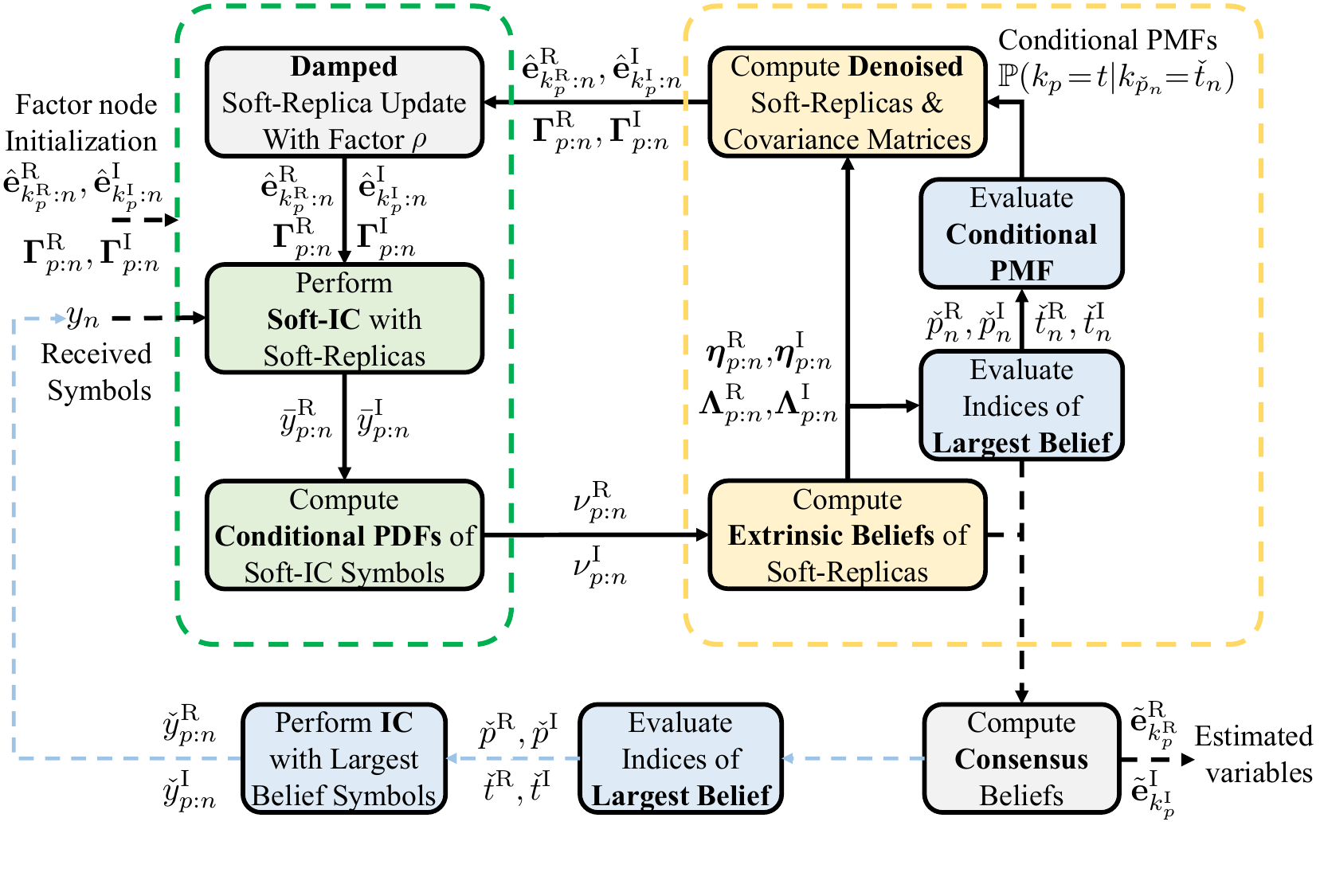}
\vspace{-5.5ex}
\caption{Schematic diagram of the enhanced \ac{UVD}-\ac{GaBP} detector with the conditional denoiser and successive \ac{IC} operations.}
\label{fig:schematicsV2}
\vspace{-1ex}
\end{figure}

The complete procedure, incorporating the proposed receiver-side enhancements of Sections \ref{sec:greedy_IC} and \ref{sec:conditional_denoiser}, is provided as a pseudocode in Algorithm \ref{alg:UVD-GaBP_SIC_Cond}, and as a schematic diagram in Figure \ref{fig:schematicsV2}.
Notice that for $P = 1$, the proposed enhancement is not necessary as there is only a single unit vector to evaluate, and the conditional \ac{PMF}-based denoiser is not applicable as there is no other unit vector to condition the extrinsic beliefs.
In addition, the motivation of such enhancements is not necessary for $P = 1$, as the \textit{duplicate convergence} error cannot occur with a single unit vector variable.


\section{Performance Analysis}
\label{sec:performance_analysis}

\subsection{\acs{BER} Performance}
\label{sec:BER_performance}

To analyze the performance of the proposed \ac{UVD}-\ac{GaBP} algorithm and the enhancement modules in Sec. \ref{sec:enhanced_decoder}, we perform simulations for different system sizes characterized by $N_T$ transmit antennas, $N_R$ receive antennas, and $P$ pilot symbols from the $M$-ary constellation, {\color{black}including truly massive systems with large $N_T$ and $N_R$.
Specifically, three algorithmic cases as summarized in Table \ref{tab:sim_cases}, are selected} to capture and highlight the improvements introduced by each of the three proposed enhancement modules, namely: Algorithm \ref{alg:UVD-GaBP} as the baseline \ac{UVD}-\ac{GaBP} decoder; Algorithm \ref{alg:UVD-GaBP_SIC_Cond} with conditional denoiser but without successive \ac{IC}; and finally the full Algorithm \ref{alg:UVD-GaBP_SIC_Cond} with both the conditional denoiser and successive \ac{IC}.
Note that the optimized constellations are employed in all cases considered.

To serve as a reference, the results are compared against the \ac{MFB}, which corresponds to the performance of a Genie-aided system in which  the soft-replicas of the \ac{GaBP} iterations are initialized with the actual (true) transmitted activation vectors.

Our first set of results, depicted in Figure \ref{fig:BER_NT32_perP}, compares the \ac{BER} performances of the proposed \ac{UVD}-\ac{GaBP} decoder for all cases in Table \ref{tab:sim_cases} and the corresponding \ac{MFB}, in \ac{MIMO} systems of size $32\times 32$ employing various numbers of symbols $P$.

\begin{table}[H]
\centering
\caption{Selected cases for the numerical evaluation.}
\label{tab:sim_cases}
\vspace{-1ex}
\begin{tabular}{|c||c|c|c|}
\hline
& Alg. 1 & Alg. 2 (no \ac{IC}) & Alg. 2 \\ 
\hline \hline
Opt. Constellation (Sec. \ref{sec:pilot_constellation_design}) & $\checkmark$ & $\checkmark$ & $\checkmark$ \\ \hline
Cond. Denoising (Sec. \ref{sec:conditional_denoiser}) &  & $\checkmark$ &  $\checkmark$ \\ \hline
Successive \ac{IC} (Sec. \ref{sec:greedy_IC}) &  &  & $\checkmark$ \\ \hline\hline
\end{tabular}
\end{table}

\begin{figure}[t]
\includegraphics[width=1\columnwidth]{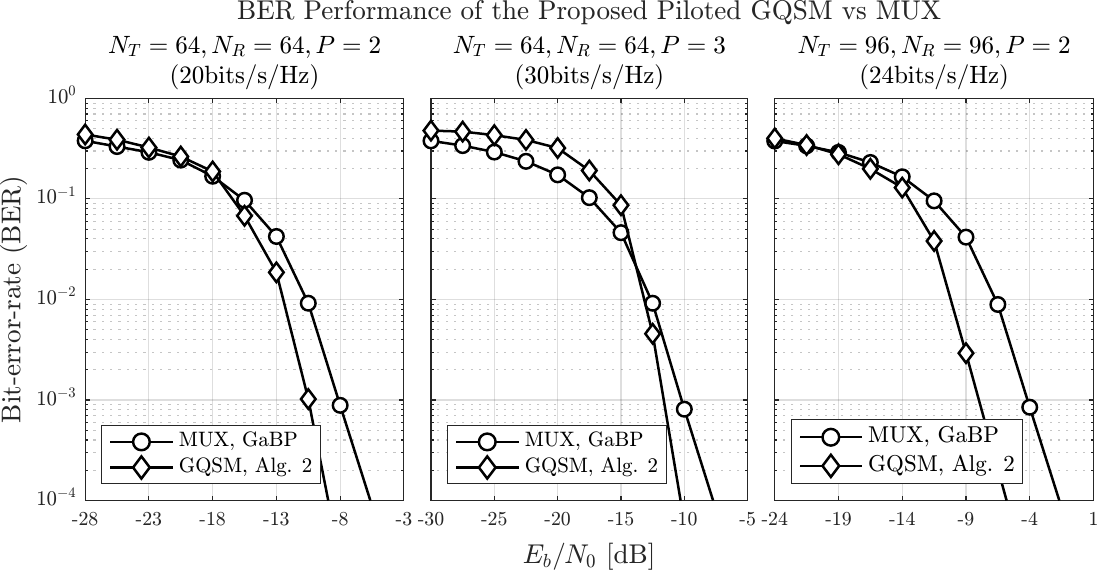}
\vspace{-3ex}
\caption{The \ac{BER} performance comparison of piloted \ac{GQSM}-\ac{ISAC} with the proposed decoder of Algorithm \ref{alg:UVD-GaBP_SIC_Cond} and piloted MUX-\ac{ISAC} with linear \ac{GaBP} detection.}
\label{fig:BER_Comp}
\vspace{-1.5ex}
\end{figure}

{\color{black} The \ac{BER} plots against the \ac{EbN0} capture the effects of the duplicate convergence of unit vectors as identified in  Section \ref{sec:enhanced_decoder}, highlighting it as a source of errors, which of course does not occur for $P = 1$ but can be substantial for $P>1$.
The results show, however, that this error is successfully mitigated by the improvements described in Sections \ref{sec:conditional_denoiser} and \ref{sec:greedy_IC}.

In addition, the plots also exhibit error-floors for $P > 1$, which still remain while significantly lowered by the proposed enhancement modules.
Such error-floor is a convergent phenomenon of \ac{MP} algorithms in non-ideal high \ac{SNR} regimes \cite{Schlegel_TIT10,Knoll_2019}, also enlarged when the number of observation nodes is small, i.e., $N_R < N_T$, whose mitigation has been extensively studied via various methods, including adaptive \ac{MP} updates and post-processing \cite{Takahashi_TC19,Zhang_GLOBECOM08}.
However, such extra modifications to further reduce the error-floors are not specific to the \ac{IM} structure and its detection, and therefore have been considered out of scope of this article.

\begin{figure}[t]
\centering
\captionsetup[subfloat]{labelfont=small,textfont=small}
\subfloat[$P = 1$ and $P = 2$.]{\includegraphics[width=0.99\columnwidth]{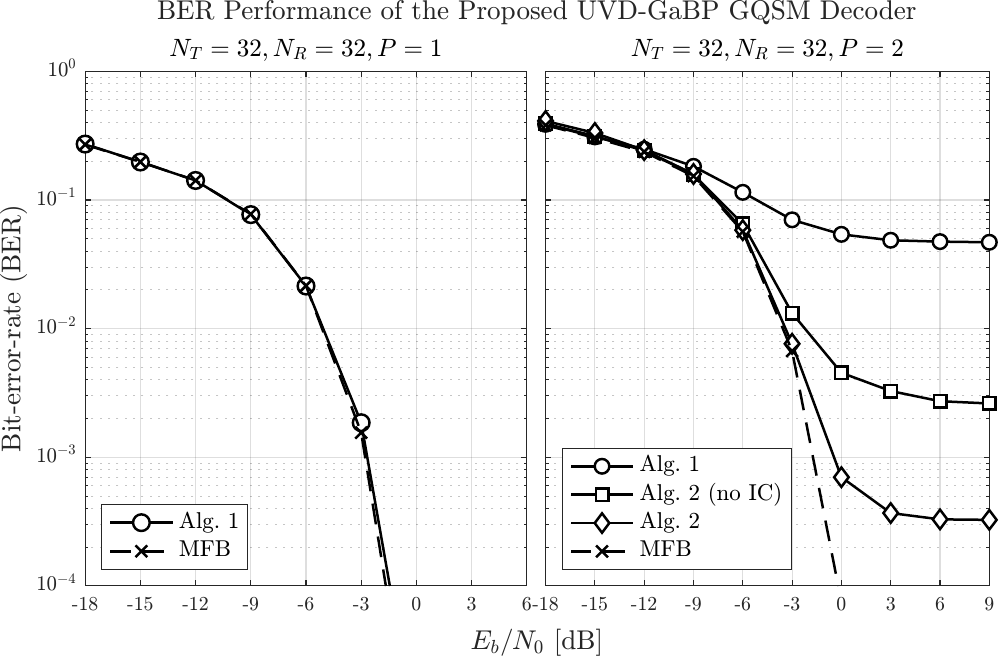}%
\label{fig:BER_NT32_P12}}
\vspace{-1ex}
\captionsetup[subfloat]{labelfont=small,textfont=small}
\subfloat[$P = 3$ and $P = 4$.]{\includegraphics[width=0.99\columnwidth]{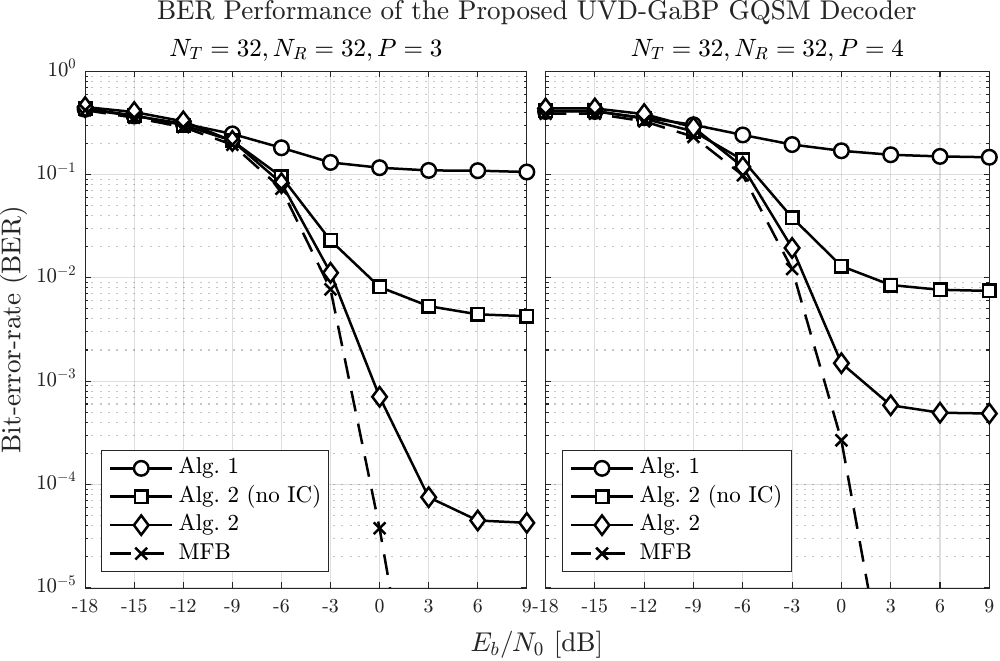}%
\label{fig:BER_NT32_P34}}
\vspace{-1ex}
\caption{The \ac{BER} performance comparison for the different cases of the proposed \ac{UVD}-\ac{GaBP} decoder, with system size parameters $N_T = 32$, $N_R = 32$, $M = 4$.}
\label{fig:BER_NT32_perP}
\vspace{-2ex}
\end{figure}

Furthermore, it is found that sufficiently large \ac{MIMO} system sizes can sustain larger $P$ values with significantly reduced error-floors.
In order to demonstrate this, we offer our next set of results, given in Figure \ref{fig:BER_Comp}, which show the performances achieved by Algorithm \ref{alg:UVD-GaBP_SIC_Cond} in truly massive system scales of $64\times 64$ and $96\times 96$.
We remark that simulation results for \ac{GQSM} systems of such sizes have never been shown before, precisely due to the prohibitive complexity even with the \ac{SotA} \ac{GQSM} detection methods \cite{Rou_TWC22,An_TVT22}.
%

Notice that Figure \ref{fig:BER_Comp} also compares the performance of the proposed piloted \ac{GQSM} scheme against a conventional multiplexed (MUX) \ac{MIMO} scheme, which in order to enable a fair comparison between the two (proposed and conventional) approaches, the total power and transmission rate (in bits/s/Hz) of both systems have been fixed to identical values, and employ the conventional linear \ac{GaBP} \cite{Som_ITW10,Takahashi_TC19} in the detection of the MUX scheme.
It can be seen that, piloted \ac{GQSM} in massive systems, enabled by the proposed \ac{UVD}-\ac{GaBP} detector, outperforms a conventional \ac{MIMO}-MUX system under equal power and resource constraints, without pilots.

In light of the latter finding, one highly promising application and motivation towards the proposed massive \ac{IM} framework can be highlighted, which is the efficient incorporation of \ac{ISAC} via the exploitation of the pilot symbols.
Specifically, as the \ac{IM} framework only embeds information in the index domain and not in the symbols themselves, the sensing functionality can be performed with the known (pilot) symbols via sensing methods available in the literature \cite{Gaudio_TWC20, Hua_TWC24, Rou_TWC24, Ranasinghe_ICASSP24,Ranasinghe_Arxiv24,Rou_SPM24} without interference from unknown communication symbols, which is known to be a significant challenge in \ac{ISAC} systems in which decoding errors made on communication symbol estimates may deteriorate sensing performance \cite{Mi_TC17,Peng_ICWCSP13}, while channel estimation errors made over pilots may impact on \ac{BER} performance \cite{Sugiura_SPL12}.
}

\vspace{-2ex}

\subsection{Decoding Complexity}
\label{sec:complexity_analysis}

In this section, the decoding complexities of the \ac{SotA} methods, the proposed \ac{UVD}-\ac{GaBP} decoder of Sec. \ref{sec:proposed_decoder}, and the enhancement modules in Section \ref{sec:enhanced_decoder} are derived, highlighting the significantly reduced complexity, that is completely independent of the prohibitive combinatorial coefficient. 
Specifically, the computational complexity $\mathcal{C}$ is evaluated in terms of the number of \acp{FLOP}, which represents the cost of a single arithmetic operation of a real-valued scalar, and the complexity order $\mathcal{O}(\,\cdot\,)$ representing the leading coefficient order of each system parameter variable.

\begin{figure*}[b]
    \centering
    \vspace{-2ex}
    \hrulefill
    \vspace{-1ex}
    \small
    \setcounter{equation}{55}
    \begin{subequations}
    \begin{align}
    \mathcal{C}_\mathrm{ML} & \!\triangleq\! \mathsmaller{\binom{N_T}{P}}^{\!2} \!\!\cdot\,\! (8N_R N_T + 4N_R), 
    \label{eq:comp_ML}
    \\
    \mathcal{C}_\mathrm{IQ} & \!\triangleq\! \tau \big(2N_R [\mathsmaller{\binom{N_T}{P}}(15N_T^2 + 15N_T + 4) +3N_T^2 - 2] + 4N_T^2\big) + 2N_R(N_T^2 + 2N_T + 1) \!+\! \mathsmaller{\binom{N_T}{P}}(6N_T^2 + 9N_T + 2) \!+\! 5N_T^2,
    \label{eq:comp_IQ}
    \\
    \mathcal{C}_\mathrm{UVD} &\!\triangleq\! \tau \!\cdot 4N_RP [ 2P(6N_T^2 + 3N_T - 1) + 6N_T^2 + 8N_TN_R + 4N_T + 4N_R + 1 ] + P(8N_T N_R + 4N_R + 10N_T + 4), 
    \label{eq:comp_UVD}
    \\
    \!\!\!\!\!\mathcal{C}_\mathrm{E\text{-}UVD} &\!\triangleq\! \tau \!\cdot 4N_RP^2 [ 2P(6N_T^2 \!+\! 3N_T \!+\! 2N_TN_R \!-\! 1) + 6N_T^2 \!+\! 8N_TN_R + 4N_T + 4N_R + 1 ] + P^2(12N_T N_R + 4N_R + 10N_T + 4). \!\!\!\! 
    \label{eq:comp_EUVD}
    \end{align}
    \label{eq:complexity_all}
    \end{subequations}
    \vspace{-1.5ex}
    \end{figure*}

\begin{table}[t]
\vspace{0.5ex}
\centering
\caption{Computational complexity orders of the \ac{SotA} \ac{GQSM} decoders the proposed \ac{UVD}-{GaBP} decoders.}
\label{tab:complexities}
\vspace{-0.5ex}
\begin{tabular}{|c|c|c|}
\hline
Algorithm & Complexity Order & Relative Complexity Order\\ 
\hline \hline
Brute-force \acs{ML} & $\mathcal{O}\big[\binom{N_T}{P}^{\!2} N_T N_R \big]$ & $\binom{N_T}{P}^{\!2}$ \\[0.5ex] \hline
\ac{IQ}-VGaBP \cite{Rou_Asilomar22_QSM} & $\mathcal{O}\big[\tau \binom{N_T}{P} N_T^2 N_R \big]$ & $\tau \binom{N_T}{P} N_T$ \\[0.5ex] \hline
Proposed - Alg. \ref{alg:UVD-GaBP} & $\mathcal{O}\big[\tau P^2 N_T^2 N_R^2  \big]$ & $\tau  P^2 N_T N_R$ \\[0.5ex] \hline Proposed - Alg. \ref{alg:UVD-GaBP_SIC_Cond} & $\mathcal{O}\big[\tau P^3 N_T^2 N_R^2  \big]$ & $\tau P^3 N_T N_R$ \\[0.5ex] \hline
\hline
\end{tabular}
\vspace{-1ex}
\end{table}

{\color{black} In light of the above, Table \ref{tab:complexities} summarizes the decoding complexity of four detection algorithms for piloted \ac{GQSM} in both the complexity order and relative complexity in \acp{FLOP}.

\begin{figure}[t]
\vspace{0.5ex}
\centering
\includegraphics[width=0.98\columnwidth]{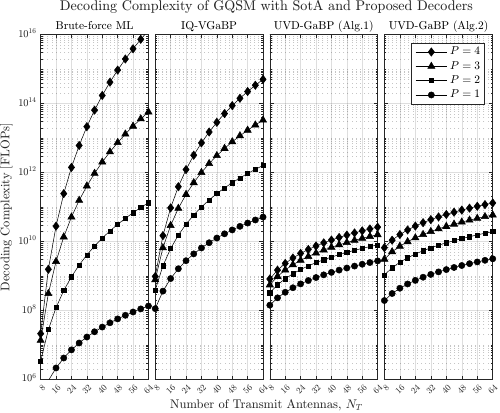}
\caption{\color{black}Decoding complexity of the considered algorithms, with varying number of transmit antennas $N_T$ and number of transmit symbols $P$, fixed $N_R = 64$, and $\tau = 100$.}
\label{fig:complex}
\vspace{-1.75ex}
\end{figure}

For the brute-force \ac{ML} detector, the likelihood metric is evaluated for all codewords of the piloted \ac{GQSM} codebook $\mathcal{X}$, and the codeword with the smallest metric is selected, \textit{i.e.,} \vspace{-2ex}
\setcounter{equation}{54}
\begin{equation}
\hat{\bm{x}}_{\mathrm{ML}} = \underset{\bm{x} \in \mathcal{X}}{\mathrm{argmin}} \;\! || \bm{y} - \bm{H}\bm{x} ||_2^2,
\label{eq:ML_metric}
\vspace{-1ex}
\end{equation}
where the required computational complexity in \acp{FLOP} is given by equation \eqref{eq:comp_ML}, which is trivially dependent on the combinatoric full codebook size.

Next, the \ac{SotA} \ac{IQ}-decoupled \ac{VGaBP} scheme of \cite{Rou_Asilomar22_QSM} is analyzed, which was shown to approach the optimal performance of the \ac{ML} decoder. 
Following \cite{Rou_Asilomar22_QSM}, the decoding complexity of the \ac{IQ}-decoupled \ac{VGaBP} is given by equation \eqref{eq:comp_IQ}, with $\tau$ number of \ac{MP} iterations, where a quadratic order complexity reduction is attained by exploiting the half-sized decoupled \ac{i.i.d.} variables of the \ac{GQSM} in the \ac{IQ} domain of equation \eqref{eq:GQSM_txvec}.

Finally, the decoding complexities of the two proposed algorithms are obtained by following each step of the proposed \ac{MP} rules elaborated in Sections \ref{sec:proposed_decoder} and \ref{sec:enhanced_decoder}, which are respectively given by equations \eqref{eq:comp_UVD} and \eqref{eq:comp_EUVD}.
}

By comparing the complexity order, it can be seen that the intractible combinatorial coefficient of the \ac{GQSM} is not inherited by the proposed \ac{UVD}-\ac{GaBP} decoders, and instead replaced with a significantly lower polynomial-order complexity order term on the system size parameters $P, N_T,$ and $N_R$.

Figure \ref{fig:complex} compares the full decoding complexity in \acp{FLOP} of the four analyzed decoders, for increasing number of transmit antennas $N_T$ and varying values of symbols $P$.
The number of receive antennas $N_R$ and the number of \ac{MP} iterations $\tau$ are fixed respectively at a reasonable value of $N_R = 64$ and $\tau = 100$, as the two parameters only introduce a linear complexity order, as opposed to $N_T$ and $P$ with more prominent effects, as can be seen in the summary of Table \ref{tab:complexities}.

For $P = 1$, the combinatorial coefficient $\binom{N_T}{P}$ will reduce to $N_T$ and hence the \ac{ML} decoder enjoys a significantly low complexity, however, for actual combinatorial \ac{IM} codebooks even with $P = 2$ or $P = 3$, the complexity of the \ac{ML} decoder and the \ac{IQ}-\ac{VGaBP} is shown to increase to an infeasible order with respect to $N_T$ and $P$, {\color{black}especially for significantly large scales of the transmitter with $N_T \geq 16$.}

On the other hand, the proposed \ac{UVD}-\ac{GaBP} while requiring a higher complexity than that of the \ac{ML} with small $N_T$ and $P$, a significant superiority can eb observed for scenarios with larger system parameters of $N_T$ and $P$, where the rate of complexity gain with respect to $N_T$ and $P$ is shown to be very low, noting the logarithmic scale of the complexity plot.

{\color{black} While already illustrated by the behavior for the smaller values of $P = 1,2,3,4$, it should also be highlighted that the complexity advantage between the non-combinatorial and combinatorial detectors becomes more significant for increasing values of $P$, which is not plotted for the sake of visualization.
Therefore, when $P$ is parametrized to yield the maximum spectral efficiency, which is $P = N_T/2$ for piloted \ac{IM} schemes and $P > N_T / 2$ for digitally modulated \ac{IM} schemes such as \ac{GQSM} \cite{Rou_TWC22,An_TVT22}, the combinatoric-free advantage of the proposed decoders will be more protuberant.}

\begin{figure}[t]
\centering
\includegraphics[width=1\columnwidth]{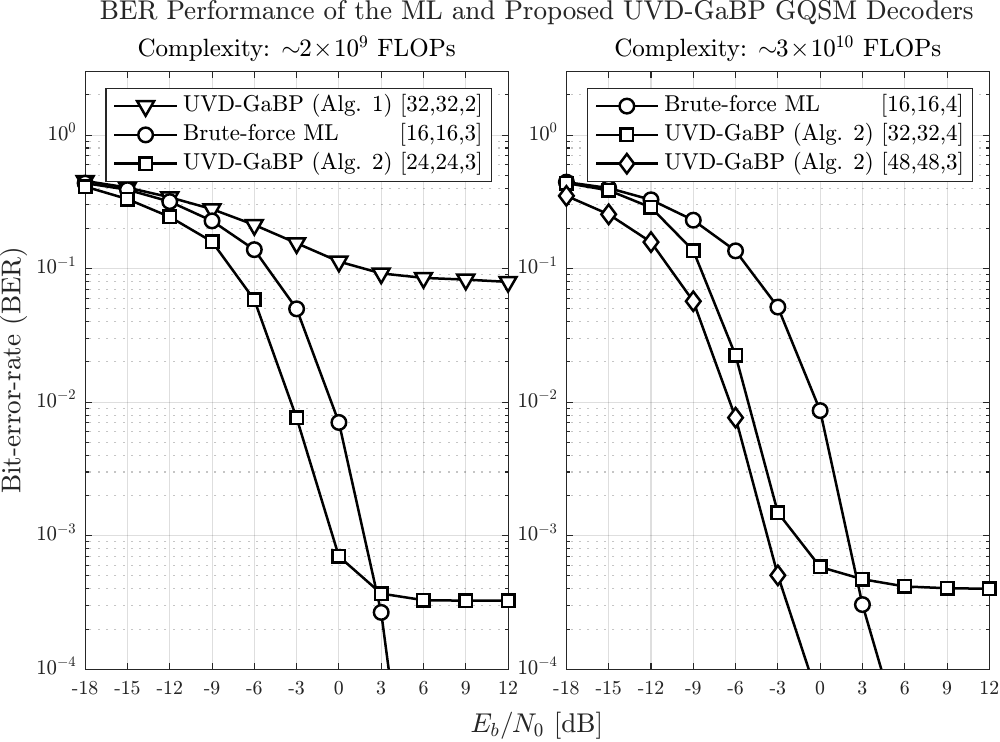}
\vspace{-3.5ex}
\caption{The \ac{BER} performance comparison for the \ac{ML} and \ac{UVD}-\ac{GaBP} decoders under varying system size parameters but the approximately same decoding complexities.}
\label{fig:comp_ber_complexity}
\end{figure}

Finally, Figure \ref{fig:comp_ber_complexity} highlights the superiority of the low-complexity \ac{UVD}-\ac{GaBP} enabled piloted \ac{GQSM} system with respect to the effective gain of the \ac{BER} performance per decoded bit, in other words, illustrates the comparison of differently parametrized \ac{MIMO} \ac{GQSM} with $N_T$ and $P$, subject to the same decoding complexity cost.

The left subfigure first compares piloted \ac{GQSM} systems requiring a decoding complexity in the order $2\!\times\!10^9$ \acp{FLOP}, with the three selected scenarios: \ac{ML} decoder with $N_T \!=\! 16$, $N_R \!=\! 16,$ $P \!=\! 3$, the enhanced \ac{UVD}-\ac{GaBP} decoder (Alg. \ref{alg:UVD-GaBP_SIC_Cond}) with $N_T = 24, N_R = 24,$ and $P = 3$, and the \ac{UVD}-\ac{GaBP} decoder (Alg. \ref{alg:UVD-GaBP}) with $N_T = 32, N_R = 32,$ and $P = 2$.
Similarly, the right subfigure compares piloted \ac{GQSM} systems requiring a decoding complexity in the order $3\!\times\!10^{10}$ \acp{FLOP}, with: \ac{ML} decoder with $N_T \!=\! 16$, $N_R \!=\! 16,$ $P \!=\! 4$, the enhanced \ac{UVD}-\ac{GaBP} decoder (Alg. \ref{alg:UVD-GaBP_SIC_Cond}) with $N_T \!=\! 32, N_R \!=\! 32,$ and $P \!=\! 4$, and also with $N_T \!=\! 48, N_R \!=\! 48,$ and $P \!=\! 3$.

In light of the decreased complexity, the supported \ac{GQSM} system of the \ac{UVD}-\ac{GaBP} decoder is able be significantly scaled especially at larger ordered, as compared to the \ac{ML} decoder.
The increased system parameters of $N_T$ and $P$ inherently imply an increased spectral efficiency of the system, improving the \ac{BER} performance with respect to the \ac{EbN0}.

The \ac{EbN0} gain is about 3$\mathrm{dB}$ in the left figure under a decoding complexity of $2\!\times\!10^9$ \acp{FLOP}, and the gain increases for higher complexity scales to 6$\mathrm{dB}$ in the right figure under a decoding complexity of $3\!\times\!10^{10}$ \acp{FLOP}.
This \ac{EbN0} gain is expected to increase with increasing complexity order, due to the combinatorial growth of the \ac{ML} decoding complexity, as can be seen in the behavior of Figure \ref{fig:complex}.

While the error floor behavior is also still present in the proposed decoders as opposed to the \ac{ML} decoders, the performance upto the error floor is superior, and the error floor is also expected to be mitigated in large systems as shown in the performance analysis of Section \ref{sec:performance_analysis}, or via post-processing techniques as previously discussed.

\section{Conclusion}
\label{sec:conclusion}
{\color{black}
We proposed a new paradigm for feasibly enabling truly massive scales of \ac{IM} schemes in expectation for the \ac{B5G}, by proposing a novel detection algorithm which is shown to achieve a decoding complexity that is completely independent of the previously prohibitive combinatorial factor. 
The proposed scheme is based on a novel \ac{UVD} formulation of piloted \ac{IM} signals which is generally applicable to many types of \ac{IM} schemes, on which the tailored \ac{GaBP} rules are designed, aided by the novel derivation of the index activation probabilities.
In addition, several enhancements to improve the performance of piloted \ac{IM} with the \ac{UVD}-\ac{GaBP} is presented, in the form of transmit symbol constellation optimization and modified \ac{MP} rules based on conditional \ac{PMF}-based denoising and successive \ac{IC}.
The effectiveness of the proposed method is verified via simulation results with the \ac{GQSM} as the exemplary technique, which is shown to outperform the classical \ac{MIMO} multiplexing scheme in terms of the \ac{BER} performance in very large \ac{MIMO} systems, in addition to showing the reduced complexity of the proposed detector by showing unprescedented \ac{GQSM} detection results for system sizes up to $96 \times 96$ transmit and receive antennas.
}


\begin{thebibliography}{10}
\providecommand{\url}[1]{#1}
\csname url@samestyle\endcsname
\providecommand{\newblock}{\relax}
\providecommand{\bibinfo}[2]{#2}
\providecommand{\BIBentrySTDinterwordspacing}{\spaceskip=0pt\relax}
\providecommand{\BIBentryALTinterwordstretchfactor}{4}
\providecommand{\BIBentryALTinterwordspacing}{\spaceskip=\fontdimen2\font plus
\BIBentryALTinterwordstretchfactor\fontdimen3\font minus
\fontdimen4\font\relax}
\providecommand{\BIBforeignlanguage}[2]{{%
\expandafter\ifx\csname l@#1\endcsname\relax
\typeout{** WARNING: IEEEtran.bst: No hyphenation pattern has been}%
\typeout{** loaded for the language `#1'. Using the pattern for}%
\typeout{** the default language instead.}%
\else
\language=\csname l@#1\endcsname
\fi
#2}}
\providecommand{\BIBdecl}{\relax}
\BIBdecl

\setlength{\baselineskip}{9.9pt}

{\color{black}
\bibitem{Rou_CAMSAP23}
H.~S. Rou, \emph{et al.}, ``Enabling
energy-efficiency in massive-{MIMO}: A scalable low-complexity decoder for
generalized quadrature spatial modulation,'' in \emph{IEEE 9th
International Workshop on Computational Advances in Multi-Sensor Adaptive
Processing}, 2023, pp. 301--305.

\bibitem{Rappaport_Access19}
T.~S. Rappaport \emph{et al.}, ``Wireless communications and applications above 100 {GHz}: Opportunities and challenges for {6G} and
beyond,'' \emph{IEEE Access}, vol.~7, pp. 78\,729--78\,757, 2019.

\bibitem{Vo_MNA22}
N.-S. Vo, T.~Q. Duong, and Z.~Sheng, ``The key trends in {B5G} technologies, services and applications,'' \emph{Mobile Networks and Applications},
vol.~27, no.~4, pp. 1716--1718, 2022.

\bibitem{Uusitalo_Access21}
M.~A. Uusitalo \emph{et al.}, ``{6G}
vision, value, use cases and technologies from european {6G} flagship project {Hexa-X},'' \emph{IEEE Access}, vol.~9, pp. 160\,004--160\,020, 2021.

\bibitem{Wang_CST23}
C.-X. Wang \emph{et al.}, ``On the road to {6G}: Visions, requirements,
key technologies and testbeds,'' \emph{IEEE Commun. Surveys \&
Tuts.}, 2023.

\bibitem{Tripathi_6G21}
S.~Tripathi, N.~V. Sabu, A.~K. Gupta, and H.~S. Dhillon, ``Millimeter-wave and
terahertz spectrum for {6G} wireless,'' in \emph{6G Mobile Wireless
Networks}.\hskip 1em plus 0.5em minus 0.4em\relax Springer, 2021, pp.
83--121.

\bibitem{Song_TTHz22}
H.-J. Song and N.~Lee, ``Terahertz communications: Challenges in the next
decade,'' \emph{IEEE Trans. THz Sci. Technol.}, vol.~12, no.~2, 2022.

\bibitem{Huo_Sensors23}
Y.~Huo \emph{et al.}, ``Technology trends for massive {MIMO} towards {6G},'' \emph{Sensors}, vol.~23, no.~13, p. 6062, 2023.

\bibitem{He_JCIN21}
H.~He \emph{et al.,} ``Cell-free massive {MIMO}
for {6G} wireless communication networks,'' \emph{Journ. Commun. Inf. Net.}, vol.~6, no.~4, pp. 321--335, 2021.

\bibitem{Zhang_Springer21}
H.~Zhang, B.~Di, L.~Song, and Z.~Han, \emph{Reconfigurable intelligent
surface-empowered 6G}.\hskip 1em plus 0.5em minus 0.4em\relax Springer, 2021.

\bibitem{Pan_CM21}
C.~Pan \emph{et al.}, \emph{et~al.}, ``Reconfigurable
intelligent surfaces for 6G systems: Principles, applications, and research
directions,'' \emph{IEEE Communications Magazine}, vol.~59, no.~6, pp.
14--20, 2021.

\bibitem{Liu_JSC22}
F.~Liu \emph{et al.},
``Integrated sensing and communications: Toward dual-functional wireless
networks for {6G} and beyond,'' \emph{IEEE Journal on Selected Areas in
Communications}, vol.~40, no.~6, pp. 1728--1767, 2022.

\bibitem{WangITJ2022}
J.~Wang \emph{et al.},
``Integrated sensing and communication: Enabling techniques, applications,
tools and data sets, standardization, and future directions,'' \emph{IEEE
Internet of Things Journal}, vol.~9, no.~23, 2022.

\bibitem{Wei_ITJ23}
Z.~Wei \emph{et al.},
``Integrated sensing and communication signals toward {5G-A} and {6G}: A
survey,'' \emph{IEEE Internet of Things Journal}, vol.~10, no.~13, pp.
11\,068--11\,092, 2023.

\bibitem{Rou_SPM24}
H.~S. Rou \emph{et al.}, ``From Orthogonal Time-Frequency Space to Affine Frequency-Division Multiplexing: A comparative study of next-generation waveforms for integrated sensing and communications in doubly dispersive channels," \emph{in IEEE Signal Processing Magazine [Special Issue on Signal Processing for the Integrated Sensing and Communications Revolution]}, vol. 41, no. 5, pp. 71-86, Sept. 2024.


\bibitem{Mandloi_B5G21}
M.~Mandloi, A.~Datta, and V.~Bhatia, ``Index modulation techniques for {5G} and
beyond wireless systems,'' \emph{5G and Beyond Wireless Systems: PHY Layer
Perspective}, pp. 63--83, 2021.


\bibitem{Mao_CST19}
T.~Mao, Q.~Wang, Z.~Wang, and S.~Chen, ``Novel index modulation techniques: A
survey,'' \emph{IEEE Communications Surveys and Tutorials}, vol.~21, no.~1,
pp. 315--348, 2019.

\bibitem{Ishikawa_CST18}
N.~Ishikawa \emph{et al.,} ``50 years of permutation, spatial and index modulation: From classic RF to visible light communications and data storage," \emph{IEEE Commun. Surv. Tuts.}, vol.~20, no.~3, pp.~1905--1938, 2018.

\bibitem{Cheng_WC18}
X.~Cheng \emph{et al.,} ``Index modulation for {5G}: Striving
to do more with less,'' \emph{IEEE Wireless Communications}, vol.~25, no.~2,
pp. 126--132, 2018.

\bibitem{Sugiura_Access17}
S.~Sugiura, T.~Ishihara, and M.~Nakao, ``State-of-the-art design of index modulation in the space, time, and frequency domains: Benefits and
fundamental limitations,'' \emph{IEEE Access}, vol.~5, pp. 21\,774--21\,790,
2017.

\bibitem{Wen_TSP15}
M.~Wen \emph{et al.,} ``On the achievable rate of OFDM with index modulation", \emph{IEEE Trans. Sig. Proc.}, vol.~64, no.~8, pp.~1919--1932, 2015.

\bibitem{Wen_TC17}
M.~Wen \emph{et al.,} ``Multiple-mode orthogonal frequency division multiplexing with index modulation, \emph{IEEE Trans. Commun.}, vol.~64, no.~9, pp.~3892--3906, 2017.

\bibitem{Li_TWC22}
J.~Li \emph{et al.,} ``Composite multiple-mode orthogonal frequency division multiplexing with index modulation," \emph{IEEE Trans. Wireless. Commun.}, vol.~22, no.~6, pp.~3748--3761, 2022.

\bibitem{Li_WC20}
Q.~Li \emph{et al.}, ``Subcarrier
index modulation for future wireless networks: Principles, applications, and
challenges,'' \emph{IEEE Wireless Communications}, vol.~27, no.~3, pp.
118--125, 2020.

\bibitem{Mesleh_TVT08}
R. Y. Mesleh \emph{et al.}, ``Spatial
modulation,'' \emph{IEEE Trans. Veh. Technol.}, vol.~57, no.~4, pp.
2228--2241, Jul. 2008.

\bibitem{Younis_Asilomar10}
{A. Younis \emph{et al.}}, ``Generalised spatial
modulation,'' in \emph{Proc. 44th Asilomar Conf. Signals, Syst. Comput.},
2010, pp. 1498--1502.

\bibitem{Wen_JSAC19}
M.~Wen \emph{et al.,} ``A survey on spatial modulation in emerging wireless systems: Research progresses and applications," \emph{IEEE Journ. Sel. Areas. Commun.,} vol.~37, no.~9, pp.~1949--1972, 2019.

\bibitem{Rou_TWC22}
H.~S. Rou \emph{et al.}, ``Scalable
quadrature spatial modulation,'' \emph{IEEE Trans. on Wireless
Commun.}, vol.~21, no.~11, pp. 9293--9311, 2022.

\bibitem{Raafat_TWC20}
A.~Raafat, A.~Agustin, and J.~Vidal, ``Downlink multi-user massive {MIMO}
transmission using receive spatial modulation,'' \emph{IEEE Transactions on
Wireless Communications}, vol.~19, no.~10, pp. 6871--6883, 2020.

\bibitem{Li_Network23}
J.~Li \emph{et al.,} ``Index modulation multiple access for 6G communications: Principles, applications, and challenges,'' \emph{IEEE Network}, vol.~37, no.~1, pp.~52--60, 2023.

\bibitem{Sui_TC24}
Z.~Sui \emph{et al.,} ``RIS-assisted cell-free massive MIMO relying on reflection pattern modulation," \emph{IEEE Trans. Commun.,} 2024.

\bibitem{Ma_JSTSP21}
D.~Ma \emph{et al.}, ``{FRaC}: {FMCW}-based
joint radar-communications system via index modulation,'' \emph{IEEE J. Sel. Topics Signal Process.}, vol.~15, no.~6, pp. 1348--1364, 2021.

\bibitem{Gopi_TWC21}
S.~Gopi, S.~Kalyani, and L.~Hanzo, ``Intelligent reflecting surface assisted
beam index-modulation for millimeter wave communication,'' \emph{IEEE
Trans. on Wireless Commun.}, vol.~20, no.~2, pp. 983--996, 2021.

\bibitem{Rou_Asilomar24}
H.~S. Rou \emph{et al.}, ``AFDM Chirp-Permutation-Index Modulation with Quantum-Accelerated Codebook Design", \emph{2024 58th Asilomar Conference on Signals, Systems, and Computers}, 2024.


\bibitem{Shamasundar_TWC22}
B.~Shamasundar \emph{et al.}, ``On the capacity of index modulation,''
\emph{IEEE Trans. on Wireless Commun.}, vol.~21, no.~11, pp.
9114--9126, 2022.


\bibitem{Ma_TVT21}
D.~Ma \emph{et al.},
``Spatial modulation for joint radar-communications systems: Design,
analysis, and hardware prototype,'' \emph{IEEE Transactions on Vehicular
Technology}, vol.~70, no.~3, pp. 2283--2298, 2021.

\bibitem{ElMai_CAMSAP23}
A.~El-Mai \emph{et al.}, ``Efficient joint radar and communication exploiting sparsity and spatial modulation,'' in \emph{IEEE CAMSAP Workshops}, 2023.

\bibitem{Hu_PC19}
\BIBentryALTinterwordspacing
Z.~Hu, F.~Chen, Y.~Liu, S.~Liu, H.~Yu, and F.~Ji, ``Low-complexity detection
for multiple-mode ofdm with index modulation,'' \emph{Physical
Communication}, vol.~34, pp. 38--47, 2019.
\BIBentrySTDinterwordspacing

\bibitem{Wang_TVT20}
{L. Wang and Z. Chen}, ``Enhanced diversity-achieving quadrature spatial
modulation with fast decodability,'' \emph{IEEE Trans. Veh. Technol.},
vol.~69, no.~6, pp. 6165--6177, 2020.

\bibitem{An_TVT22}
J.~An \emph{et al.}, ``The achievable rate analysis of
generalized quadrature spatial modulation and a pair of low-complexity detectors,'' \emph{IEEE Trans. on Veh. Technol.}, vol.~71,
no.~5, pp. 5203--5215, 2022.

\bibitem{Rou_Asilomar22_QSM}
H.~S. Rou \emph{et al.}, ``An efficient vector-valued
belief propagation decoder for quadrature spatial modulation,'' in \emph{2022
56th Asilomar Conference on Signals, Systems, and Computers}, 2022, pp.
27--31.

\bibitem{Katla_Access20}
S.~Katla \emph{et al.}, ``Deep learning assisted detection for index modulation aided mmWave
systems,'' \emph{IEEE Access}, vol.~8, 2020.

\bibitem{Kim_WCL21}
J.~Kim, H.~Ro, and H.~Park, ``Deep learning-based detector for dual mode OFDM
with index modulation,'' \emph{IEEE Wireless Communications Letters},
vol.~10, no.~7, pp. 1562--1566, 2021.

\bibitem{Yukiyoshi_VTC24}
K.~Yukiyoshi \emph{et al.}, ``Grover adaptive search for maximum-likelihood detection of generalized
spatial modulation,'' \emph{IEEE VTC-Fall}, 2024.

\bibitem{Mesleh_TVT14}
{R. Mesleh \emph{et al.}}, ``Quadrature spatial modulation,''
\emph{IEEE Trans. Veh. Technol.}, vol.~64, no.~6, pp. 2738--2742, Jun. 2015.

\bibitem{Xiao_TC19}
L.~Xiao \emph{et al.}, ``Compressive
sensing assisted generalized quadrature spatial modulation for massive {MIMO}
systems,'' \emph{IEEE Transactions on Communications}, vol.~67, no.~7, pp.
4795--4810, 2019.

\bibitem{Ishikawa_Access21}
N.~Ishikawa, ``Quantum speedup for index modulation,'' \emph{IEEE Access},
vol.~9, pp. 111\,114--111\,124, 2021.

\bibitem{Boyd_2004}
S.~P. Boyd and L.~Vandenberghe, \emph{Convex optimization}.\hskip 1em plus
0.5em minus 0.4em\relax Cambridge university press, 2004.

\bibitem{Xiao_TVT17}
L.~Xiao  \emph{et al.},
``Efficient compressive sensing detectors for generalized spatial modulation
systems,'' \emph{IEEE Trans. on Veh. Technol.}, vol.~66, no.~2,
pp. 1284--1298, 2017.

\bibitem{Xiao_TVT19}
L.~Xiao \emph{et al.}, ``Compressive
sensing assisted generalized quadrature spatial modulation for massive {MIMO}
systems,'' \emph{IEEE Transactions on Communications}, vol.~67, no.~7, pp.
4795--4810, 2019.

\bibitem{Nguyen_TVT18}
T.~Van Hong~Nguyen, S.~Sugiura, and K.~Lee, ``Low-complexity sphere
search-based adaptive spatial modulation,'' \emph{IEEE Transactions on
Vehicular Technology}, vol.~67, no.~8, pp. 7836--7840, 2018.

\bibitem{Li_TVT22}
C.~Li \emph{et al.,} ``Quadrature spatial modulation with the fourth
order transmit diversity and low-complexity sphere decoding for large-scale MIMO systems,'' \emph{IEEE Trans. Veh. Tech.}, vol.~71, no.~8, 2022.

\bibitem{Yang_TWC16}
P.~Yang \emph{et al.}, ``Transmit precoded spatial modulation: Maximizing the minimum euclidean distance versus minimizing the bit error ratio,'' \emph{IEEE Trans. on Wireless
Commun.}, vol.~15, no.~3, pp. 2054--2068, 2016.

\bibitem{Cheng_TC18}
P.~Cheng \emph{et al.}, ``A unified precoding
scheme for generalized spatial modulation,'' \emph{IEEE Trans. on Commun.}, vol.~66, no.~6, 2018.

\bibitem{LinzBook2019}
P.~Linz, \emph{Theoretical Numerical Analysis: An Introduction to Advanced
Techniques}.\hskip 1em plus 0.5em minus 0.4em\relax Dover Publications Inc.,
2019.

\bibitem{JohnRiordanBook2002}
J.~Riordan, \emph{Introduction to Combinatorial Analysis}.\hskip 1em plus 0.5em
minus 0.4em\relax Dover Publications Inc., 2002.

\bibitem{Sablonniere_JCAM93}
P.~Sablonni{\`e}re, ``Discrete bernstein bases and hahn polynomials,''
\emph{Jounr. of Comp. and Appl. Math.}, vol.~49, no. 1-3,
pp. 233--241, 1993.

\bibitem{Gaudio_TWC20}
L.~Gaudio \emph{et al.}, ``On the effectiveness of
{OTFS} for joint radar parameter estimation and communication,'' \emph{IEEE
Transactions on Wireless Communications}, vol.~19, no.~9, pp. 5951--5965,
2020.

\bibitem{Ranasinghe_ICASSP24}
K. R. R. Ranasinghe, H. S. Rou and G. T. F. de Abreu, ``Fast and Efficient Sequential Radar Parameter Estimation in MIMO-OTFS Systems," \emph{IEEE ICASSP 2024}, Seoul, Korea, Republic of, 2024, pp. 8661-8665

\bibitem{Ranasinghe_Arxiv24}
K. R. R. Ranasinghe \emph{et al.,} ``Joint Channel, Data and Radar Parameter Estimation for AFDM Systems in Doubly-Dispersive Channels," \emph{arXiv preprint arXiv:2405.16945}, 2024.

\bibitem{Rou_Asilomar22_JCAS}
H.~S. Rou \emph{et al.}, ``Asymmetric bilinear
inference for joint communications and environment sensing,'' in \emph{2022
56th Asilomar Conference on Signals, Systems, and Computers}, 2022, pp.
1111--1115.

\bibitem{Rou_TWC24}
H.~S. Rou, G.~T.~F.~de~Abreu, D.~Gonz{\'{a}}lez~G., and O.~Gonsa,
``Integrated sensing and communications for {3D} object imaging via bilinear
inference,'' \emph{IEEE Trans. on Wireless Commun.}, 2024.

\bibitem{Parker_TSP14}
J.~T. Parker, P.~Schniter, and V.~Cevher, ``Bilinear generalized approximate
message passing--part {I}: Derivation,'' \emph{IEEE Trans. Signal Process.},
vol.~62, no.~22, pp. 5839--5853, 2014.

\bibitem{Ito_TC23}
K.~Ito \emph{et al.}, ``Bilinear gaussian belief
propagation for massive MIMO detection with non-orthogonal pilots,''
\emph{IEEE Trans. on Commun.}, 2023.

\bibitem{Som_ITW10}
{P. Som \emph{et al.}}, ``Improved large-{MIMO}
detection based on damped belief propagation,'' in \emph{IEEE Inf. Theory Workshop on Inf. Theory}, 2010.

\bibitem{Su_TSP15}
Q.~Su and Y.-C. Wu, ``On convergence conditions of gaussian belief
propagation,'' \emph{IEEE Trans. on Sig. Proc.}, vol.~63, no.~5, 2015.


\bibitem{Schlegel_TIT10}
C.~Schlegel and S.~Zhang, "On the dynamics of the error floor behavior in (regular) LDPC codes," \emph{IEEE Transactions on Information Theory}, vol.~56, no.~7, pp.~3248--3264, 2010.

\bibitem{Knoll_2019}
C. Knoll, ``Understanding the Behavior of Belief Propagation: Convergence Properties, Approximation Quality, and Solution Space Analysis", \emph{arXiv preprint arXiv:2209.05464}, 2019.

\bibitem{Takahashi_TC19}
T. Takahashi, S. Ibi and S. Sampei, ``Design of Adaptively Scaled Belief in Multi-Dimensional Signal Detection for Higher-Order Modulation," \emph{IEEE Trans. Commun.}, vol. 67, no. 3, pp. 1986-2001, 2019.

\bibitem{Zhang_GLOBECOM08}
Z. Zhang \emph{et al.,} ``Lowering LDPC Error Floors by Postprocessing," \emph{IEEE GLOBECOM}, New Orleans, LA, USA, 2008, pp. 1-6.


\bibitem{Hua_TWC24}
M.~Hua \emph{et al.,} ``Integrated sensing and communication: Joint pilot and transmission design," \emph{IEEE Trans. Wireless Commun.}, 2024.


\bibitem{Mi_TC17}
D.~Mi \emph{et al.}, ``Massive MIMO
performance with imperfect channel reciprocity and channel estimation error,'' \emph{IEEE Transactions on Communications}, vol.~65, no.~9, pp.
3734--3749, 2017.

\bibitem{Peng_ICWCSP13}
P.~Xu, J.~Wang, and J.~Wang, ``Effect of pilot contamination on channel estimation in massive MIMO systems,'' in \emph{International Conference on Wireless Communications and Signal Processing}, 2013.

\bibitem{Sugiura_SPL12}
S.~Sugiura and L.~Hanzo, ``Effects of channel estimation on spatial
modulation,'' \emph{IEEE Sig. Proc. Let.}, vol.~19, no.~12, 2012.

}

\end{thebibliography}
\end{document}